\documentclass[sigplan,nonacm]{acmart}

\usepackage{graphicx}   
\usepackage{booktabs}   
\usepackage{listings}   
\usepackage{algorithm}
\usepackage{algpseudocode}
\usepackage{float}      

\definecolor{codekw}{HTML}{1F4E79}
\definecolor{codestr}{HTML}{2E6B2E}
\definecolor{codecom}{HTML}{6B7684}
\definecolor{treebg}{HTML}{E4F4E9}
\definecolor{amdblue}{HTML}{1F5C99}
\newcommand{\amd}[1]{{\color{amdblue}(#1)}}

\lstdefinestyle{cpp}{
  language=C++,
  basicstyle=\scriptsize\ttfamily,
  keywordstyle=\color{codekw}\bfseries,
  stringstyle=\color{codestr},
  commentstyle=\color{codecom}\itshape,
  morekeywords={SmallVector,Value,Location,unsigned,auto,const,std},
  showstringspaces=false,
  columns=fullflexible,
  keepspaces=true,
  aboveskip=0pt,
  belowskip=0pt,
}

\lstdefinestyle{python}{
  language=Python,
  basicstyle=\footnotesize\ttfamily,
  keywordstyle=\color{codekw}\bfseries,
  stringstyle=\color{codestr},
  commentstyle=\color{codecom}\itshape,
  emph={triton,tl,autotune,jit,Config,constexpr},
  emphstyle=\color{codekw},
  showstringspaces=false,
  columns=fullflexible,
  keepspaces=true,
  aboveskip=0pt,
  belowskip=0pt,
}

\newcommand{\dt}[1]{\mathsf{#1}}
\newcommand{\fpadd}[1]{\mathbin{\oplus_{\dt{#1}}}}
\newcommand{\fpsum}[1]{\bigoplus\nolimits^{\dt{#1}}}

\newcommand{\tcompile}{\texttt{torch.\allowbreak compile}}

\acmConference[Conf '26]{The ACM Conference}{June 2026}{City, Country}
\acmYear{2026}

\setcopyright{none}
\renewcommand\footnotetextcopyrightpermission[1]{}

\begin{document}

\title{Taming Bitwise Behavior in GPU Kernels with Tensor Core}
\subtitle{Black-Box Reconstruction, Compiler Enforcement, and Static Verification}

\author{Ziteng Yang}
\affiliation{%
  \institution{Georgia Institute of Technology}
  \city{Atlanta}
  \state{Georgia}
  \country{USA}
}
\email{ziteng.yang@gatech.edu}

\author{Nicholas J. Riasanovsky}
\affiliation{%
  \institution{Meta}
  \city{Menlo Park}
  \state{California}
  \country{USA}
}
\email{njriasan@meta.com}

\author{Warren Deng}
\affiliation{%
  \institution{Meta}
  \city{Menlo Park}
  \state{California}
  \country{USA}
}
\email{warrdeng@meta.com}

\author{Vivek Sarkar}
\affiliation{%
  \institution{Georgia Institute of Technology}
  \city{Atlanta}
  \state{Georgia}
  \country{USA}
}
\email{vsarkar@gatech.edu}

\renewcommand{\shortauthors}{Yang et al.}

\begin{abstract}
  Determinism and numerical reproducibility are increasingly required of the GPU
  kernels under machine learning systems, yet two deterministic implementations
  of one kernel differ bit for bit. The result is decided by the floating-point
  (FP) reduction order above all, and by partial-sum precision, multiply-add
  fusion and where the rounding falls: written by hand, left to the compiler by
  a block-level language like Triton, or owned by a closed library like cuBLAS
  or rocBLAS. The tile shape chosen for speed therefore chooses the arithmetic,
  and a request stops being batch invariant. Holding the order still costs
  performance, reported at up to 20\%. Also, an autotuner searching hundreds of
  configurations cannot tell which of them agree bit for bit.

  In this work we characterize what decides the bitwise behavior of GPU kernels,
  for reductions and General Matrix Multiply (GEMM). \emph{i)} We give a
  descriptor that records the parameters fixing a GEMM's reduction order, such as
  where the K axis is cut in a split-K GEMM. On that basis we give the first
  black-box reconstruction of a closed source library's arithmetic towards
  bit-level correctness, on NVIDIA cuBLAS, and rebuild it as a bitwise equivalent
  GEMM family in Triton, a block-level GPU language. It matches cuBLAS bit for
  bit at a 100\% rate on Blackwell and Hopper, and with the epilogue fused from 
  realistic LLM shape, 
  the performance matches or even exceed
  \tcompile{}. \emph{ii)} Lowering to backend, we
  enforce balanced tree reduction in Triton's compiler, with a data-layout
  optimization that brings 19 of 27 kernels on GB300 and H100 within 10\% of the
  free-order mode and 5 even exceed. \emph{iii)} We build the first sound static equivalence
  checkers to decide bitwise equivalence between compiled GPU kernels, and the
  first to partition two vendors' instruction sets, NVIDIA PTX and AMD GCN with integration into Triton's autotuner as a static pruning predicate, so a search runs
  inside a single bitwise-equivalence class.

\end{abstract}

\ccsdesc[500]{Software and its engineering}

\keywords{bitwise reproducibility, numerical determinism, floating-point
  reduction order, compilers, compiler backends, code generation, program
  equivalence, static equivalence checking, black-box reconstruction, GPU
  kernels, GEMM, tensor compilers, Triton, autotuning, machine learning systems}

\maketitle

\section{Introduction}
\label{sec:intro}


Numerical determinism is now an important need for the systems that train and
serve large models: the same computation, the same bits, whenever and wherever
it runs. Without determinism during numerical computation, a model may answer differently
when asked the same question twice with the same weights and the same seed.
A kernel is \emph{batch invariant} when one request's output is the same whatever else
shares its batch, and the kernels serving these models are not. 
Serving a request alongside others changes how many rows a kernel reduces at once, which
changes the algorithm its partial sums are combined by, giving a different
floating-point reduction order, while floating-point addition
is not associative, so the bits move~\cite{he2025defeating}.   
Measured across batch size, GPU count and GPU generation on one 7B model in bfloat16, the
difference is worth up to nine percentage points of accuracy and generations
that differ in length by thousands of tokens~\cite{llmnondeterminism2025}.

The same instability reaches training. Two runs of one configuration, differing only in
effects this small, end at meaningfully different models, and a retrained model
changes its mind about individual examples it used to get
right~\cite{summers2021nondeterminism,bhojanapalli2021reproducibility}.
Reinforcement learning is worse than either, because it runs two engines at
once: a rollout engine generates, a training engine scores, and when the two
disagree about a token's log-probability the objective being optimized is no
longer the one that was written down~\cite{fp16mismatch}.


Floating-point accumulation order dominates~\cite{shanmugavelu2024fpna}. The precision the partial sums are held at, whether a multiply and an add were contracted into one
instruction (FMA fusion), and where the rounding to the output type falls all move the result
as well. At different infrastructure stack levels, such problems are addressed with heavy effort:
\emph{i)} \textbf{Compute in higher precision.} A fixed-point accumulator of
2098 bits for a sum, 4288 for a dot product, holds every fp64 magnitude, so
every addition is exact and one rounding closes
it~\cite{demmel2013reproducible,collange2015reduction,ahrens2020reproblas}. The
\textbf{Ozaki scheme} reaches matrix multiplication the other way, splitting
each input into int8 pieces a GPU's integer matrix unit multiplies exactly and
accumulates in int32~\cite{ozaki2025}. Both cost several times the work.
\emph{ii)} \textbf{Set library mode.} To give callers the property without
asking them to change an algorithm, production libraries attach conditions to
it. Intel's \textbf{oneMKL} returns bit-identical results run to run only while the
executable, the instruction-set code path and the thread count all stay fixed,
and warns that pinning the code path can more than halve its speed; PyTorch's
deterministic mode substitutes deterministic operators, raises an error where
none exists, and scopes its guarantee to one release on one
platform~\cite{mkl_cnr,pytorch_determinism}. The price is the conditions
themselves, which the caller must hold.
\emph{iii)} \textbf{Rewrite the GPU kernels.} Nondeterministic LLM inference has been
traced to kernels whose
reduction order moves with batch size, and rewriting them to fix the order costs
about 20\% of matrix-multiply throughput against
\textbf{cuBLAS}~\cite{he2025defeating}; \textbf{RepDL} enforces correct rounding and order
invariance to keep training and inference bit-identical across
machines~\cite{repdl2025}; \textbf{LayerCast} holds weights in sixteen bits while
performing every computation in fp32, paying in bandwidth rather than in kernel
structure~\cite{llmnondeterminism2025}. \textbf{DeepSeek-V4}~\cite{deepseekv4} makes a frontier training stack bitwise
batch-invariant and deterministic end to end.
\emph{a)} cuBLAS picks its kernel from the problem dimensions behind a closed
cost model, so the arithmetic moves with the batch; a matrix-multiply library of
their own replaces it, and split-K, taken only at small batch sizes and so
batch-dependent in the same way, is dropped wherever a kernel can do without it.
\emph{b)} Atomic adds land in thread arrival order, so the backward pass gives
each streaming multiprocessor its own buffer and sums those in a fixed order.
\emph{c)} Its compiler turns fast-math off by default, aligns its lowering with
the reference CUDA toolchain so no transformation moves a bit, and calls an SMT
solver for the integer facts its layout inference rests on.
Their matrix multiplication is reported to match or surpass standard split-K in
most major scenarios. \textit{But none of them reproduces the discipline a closed library follows.} 
\emph{iv)} \textbf{Align two engines.} A reinforcement-learning rollout engine
and training engine that disagree on a token's log-probability shift the
objective being optimized; one answer moves the whole pipeline from bf16 to
fp16, trading dynamic range for the rounding headroom that keeps the two
consistent~\cite{fp16mismatch}.

Holding the bits still is a challenge at different layers:

\textbf{\textit{In the training algorithm,}} a policy update weighs each token
by the ratio of two probabilities, so when the engine that generated it and the
engine that scores it disagree the ratio is taken between two different models
rather than between two policies of one~\cite{fp16mismatch}.
\textbf{\textit{In the ML system infrastructure,}} a stack is assembled from
independently built parts, and one that accumulates differently breaks the
agreement for all of them. \textbf{\textit{Below the kernel,}} the vendor's
assembler may reassociate an accumulation it judges safe to move, and where no
machine code was shipped the driver assembles at load time, so a driver update
can change bit-level semantics the program did not.

\textbf{\textit{In the GPU kernel (scope of this work).}} 
Various obstacles exist to reaching numerical consistency:
\emph{i)} A vendor library will not fully disclose what arithmetic it performed, so a
kernel written elsewhere has difficulty matching it.
\emph{ii)} Restricting the computation order 
reduces the flexibility of optimization towards parallelism. 
PyTorch's deterministic mode stops tuning a reduction kernel's
configurations, and when its full search is enabled it takes the vendor's matrix
multiply instead of a generated one~\cite{pytorch_determinism}. 
Where agreement cannot be established, speed is what gets given up.
\emph{iii)} One source compiled twice by one compiler under different parameters
may or may not agree, and an autotuner~\cite{tvm2018,ansor2020} searching
hundreds of configurations for speed has no way to do equivalence partition.

This paper makes four contributions.

\textit{A mechanistic account of bitwise behavior.} 
We show that FP reduction order is the dominant factor that decides the numeric bits.
We give a theoretical characterization of the factors deciding the FP reduction
order of reduction kernels and General Matrix Multiply (GEMM) kernels.
We give a formal descriptor
for GEMM kernels, \emph{GEMMDesc}, that describes the full bitwise behavior of
the most commonly used GEMM variants. It records the parameters that decide the
reduction order, for example where to cut the K axis in a split-K GEMM, and
holds nothing that is only a speed knob. Two machines handed the same descriptor
owe each other the same bits.

\textit{The first Triton GEMM family that is bit-identical to a
closed-source vendor library.} NVIDIA's cuBLAS~\cite{cublas_docs} is a black-box GPU kernel
library that only provides incomplete computation information that decides the
FP computation order.
We applied the mechanism above, conducted a set of offline experiments
(runtime profiling, numerical inference, and others), and give the first
black-box reconstruction of such a library towards bit-level correctness: a
Triton GEMM family over fp16, bf16 and fp8 e4m3 that reaches 100\% bit-matches
with cuBLAS GEMM on GB300, GB200 and H100. As an extra bonus, we discovered a logical bug in
cuBLAS GEMM during that offline work
(Appendix~\ref{app:cublasbug}).\footnote{The 100\% rate excludes this bug.}


\textit{First Balanced Tree reduction in and layout optimization in Triton compiler} We
implement compiler enforcement for balanced tree reduction (``inner tree'' mode) in the
compiler backend of \textbf{Triton}~\cite{triton2019} (none before), the block-level GPU
kernel language. We analyze and implement the data-layout optimization
opportunity that the enforcement opens, bringing 19 of 27 kernels across GB300 and
H100 within 10\% of the free-order mode, and even past it on five of the ten on
H100.

\textit{The first Static \& sound equivalence checkers for compiled Triton kernels 
and integration in autotuner.} 
We implemented the first static checker that decides the bitwise equivalence
of two NVIDIA PTX kernel assembly, sound by construction, and the first
symmetric one for a second vendor's instruction set, AMD GCN.
It acts both as a static equivalence partitioner during the autotuning stage of
a GPU kernel and as a verification of the mechanism we claim. We incorparate it into
\textbf{Triton}'s autotuner and achieved static pruning before 
the actual pruning happens for the first time.
The checker is validated to be robustly \textbf{sound} across a large kernel
suite, including fused kernels generated by PyTorch
Inductor~\cite{pytorch2_2024} and Flash Attention
kernels~\cite{flashattention2022}. It recovers the exact partition on the GEMM
family and on Inductor's fused kernels, and comes within a factor 2.5 on
reductions and normalizations.


\section{Background}
\label{sec:background}

\subsection{GPU kernels and the programming model}
\label{sec:kernels}

A GPU \emph{kernel} is a function the host \emph{launches} as a \emph{grid} of
\emph{thread blocks}~\cite{cuda_prog_guide,hip_prog_model,nickolls2008cuda};
every thread runs the same body and reads an index to find its slice of the
data.
\emph{i)} A block is at most 1024 threads and goes whole onto one \emph{streaming
multiprocessor} (SM) on NVIDIA, one \emph{compute unit} (CU) on
AMD~\cite{lindholm2008tesla}, staying there until it finishes; an SM holds the
\emph{tensor cores} (AMD's \emph{matrix cores}) that execute matrix-multiply
instructions, and up to 32 blocks share one when registers and scratchpad
allow~\cite{blackwell_tuning}.
\emph{ii)} Within a block, threads are cut into fixed groups, a \emph{warp} of 32
on NVIDIA and a \emph{wavefront} of 64 on AMD's CDNA~\cite{cdna3_whitepaper},
which is the unit of instruction issue; lanes on a different branch sit an
instruction out, and since Volta~\cite{volta_whitepaper} each lane carries its
own program counter, so a value passed between lanes through memory needs an
explicit \texttt{\_\_syncwarp()}.
\emph{iii)} Above the grid, ordering across launches comes from a kernel ending,
and combining results across GPUs is a collective library's job, \textbf{NCCL} or
\textbf{RCCL}~\cite{thakur2005collectives}.

Registers are private to a thread, at most 255 of an SM's 64K. \emph{Shared
memory} (AMD's \emph{local data share}) is a scratchpad, fast on-chip memory the
program fills and empties itself, carved per resident block from the SM's
228\,KB~\cite{blackwell_tuning}. Global memory is off-chip DRAM at roughly twenty
times a shared-memory read.
Only block scope costs a barrier: \texttt{\_\_sync\allowbreak threads()} publishes a block's
shared-memory writes, a shuffle between lanes is free, and grid-wide ordering
comes from ending the kernel.

\textbf{Thread-level programming and the vendor libraries.}
CUDA C++ asks for the body of one thread and runs it across the grid, leaving
cooperation to be written by hand: which thread reads which element, when a tile
is staged into shared memory, where the barriers go. A caller can reach for a
closed-source \emph{vendor library} instead, \textbf{cuBLAS}~\cite{cublas_docs},
\textbf{cuDNN}~\cite{cudnn2014} on NVIDIA or \textbf{rocBLAS}~\cite{rocblas_docs}
on AMD, and then chooses the result while the library chooses the arithmetic that
produces it.

\textbf{Block-level programming and the compilation pipeline.}
Block-level programming asks instead for a block of threads over a tile of data
and leaves the thread detail to the compiler. \textbf{Triton}~\cite{triton2019},
\textbf{TileLang}~\cite{tilelang2026} and NVIDIA's \textbf{cuTile}~\cite{cutile}
implement it on GPUs, as do \textbf{Pallas}~\cite{pallas} on Google's
TPUs~\cite{tpu2017} and the \textbf{Neuron Kernel Interface}~\cite{nki} on AWS's
training and inference chips~\cite{trainium}. Tile operations carry no hardware
in them; the \emph{tile-level IR} fixes a \emph{layout} per tensor, saying
algebraically which lane and register hold which
element~\cite{linearlayouts2026}; an LLVM backend lowers that per thread; and
\emph{GPU assembly}, PTX or AMDGCN, is the last textual form before machine code.
The tile-level IR is where a reduction's order first becomes visible, and the
assembly is the lowest level we can read.

\textbf{Autotuning.}
Tile sizes, the pipeline depth over which a loop's loads run ahead of its
arithmetic, and warp count can be left as template parameters (\emph{knobs}). A
kernel written this way is a \emph{kernel template}, one assignment to its knobs
a \emph{configuration}, compiling under one a \emph{kernel instance}, and the
legal assignments its \emph{configuration space}, thousands for a GEMM; they are
the configurations fanning out from one template. An \emph{autotuner} picks among them by measurement
at run time, because the winner turns on occupancy, on whether the tile shape
matches the tensor core's, and on register allocation inside \texttt{ptxas}, none
of it readable off the
source~\cite{tvm2018,ansor2020}. Autotuners have become routine recently:
PyTorch's compiler generates the Triton kernels with autotuning decorators and autotunes them
~\cite{triton2019,pytorch2_2024}.

Triton spells this as an annotation on the kernel:

\begin{lstlisting}[style=python]
 @triton.autotune(configs=[
   triton.Config({'BLOCK_M': 128, 'BLOCK_K': 64},
                 num_warps=8, ...),
   triton.Config({'BLOCK_M': 64, 'BLOCK_K': 32},
                 num_warps=4, ...)],
   ..., key=['M', 'N', 'K'])
 @triton.jit
 def matmul(a, b, c, M, N, K, BLOCK_M, BLOCK_K): ...
\end{lstlisting}

The decorator compiles and times every listed configuration and caches the fastest; 
new shapes restart
the measurement. Two instances of one template compute the same mathematics, but not necessarily
the same bits.


\subsection{Related work on compiler correctness}
\label{sec:related}

\begin{figure*}[t]
  \centering
  \includegraphics[width=\textwidth]{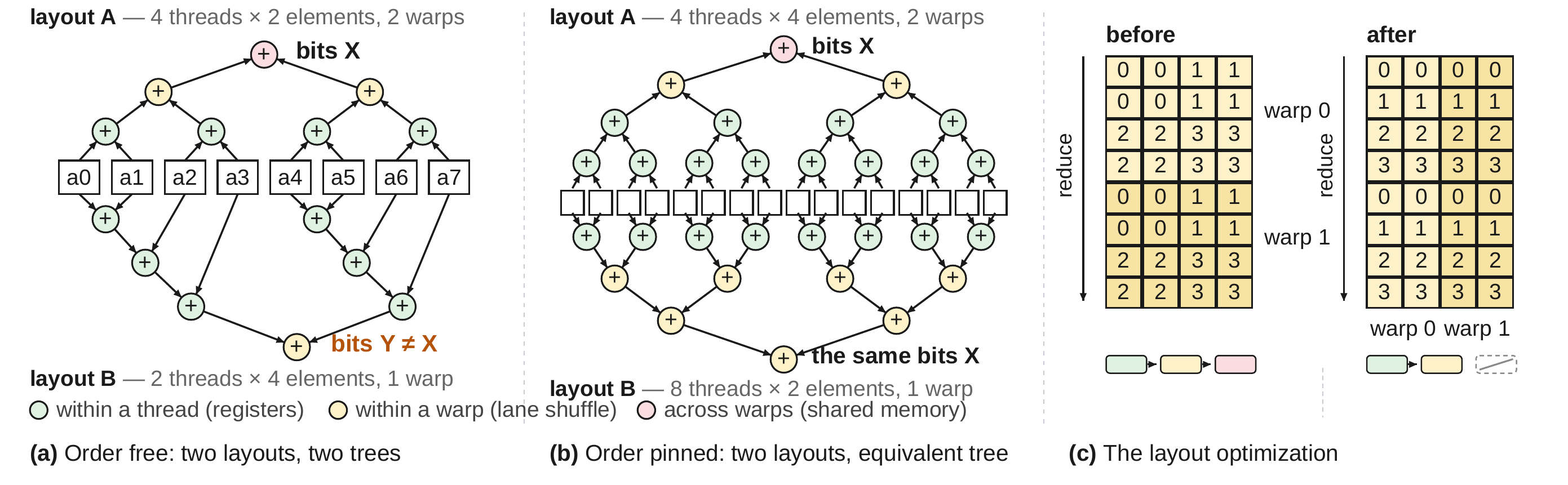}
  \caption{Pinning the reduction order, and what it buys.}
  \label{fig:inner-tree}
\end{figure*}

\textbf{Correct by construction.}
A verified compiler carries a machine-checked proof that every program it
compiles keeps its source semantics. \textbf{CompCert} is the first realistic
instance for C~\cite{compcert2009}, and its floating-point semantics are verified
in Coq (now Rocq~\cite{rocq_prover}), so IEEE-754 arithmetic survives the
transformations a compiler is otherwise tempted to
make~\cite{boldo2015fpcompile}. The same has since been done for instruction
scheduling~\cite{verifiedsched2024}, tensor-language
lowering~\cite{verifiedtensor2024}, a GPU memory consistency
model~\cite{ptxmemmodel2019} and the semantics of GPU
assembly~\cite{cudaaucoq2019}. The price is the proof effort, and a toolchain of
its own to run under.

\textbf{Translation validation.}
It drops the claim about the compiler and checks the one compilation in front of
it~\cite{pnueli1998tv}: encode both IRs and ask a solver whether the output
refines the input. \textbf{Alive2} does this for LLVM IR~\cite{alive2_2021} and
\textbf{MLIR-TV} for the multi-level IR deep-learning compilers are built
on~\cite{mlirtv2022}. \textbf{Alive-FP} brings floating point in, verifying
LLVM's floating-point and fast-math peepholes under one SMT encoding per reading
of the under-specification around signed zeros, NaNs and
infinities~\cite{alivefp2016}. The price is what the encoding admits: MLIR-TV
over-approximates floating-point arithmetic and reductions so that a
solver finish in bounded time.

\textbf{Fuzzing.}
Fuzzing proves nothing and locates real incorrectness. \textbf{Csmith} generates
random C free of undefined behavior and differentially tests compilers against
each other~\cite{csmith2011}, and \textbf{MLIRSmith} carries the idea to the
multi-level IR~\cite{mlirsmith2023}. Floating point moves the target from one
wrong compilation to two that disagree~\cite{flit2017,varity2020}. The price is
every test-based method's: finding nothing is not showing nothing is there.

None of them decides whether two compiled GPU kernels return identical
bits, let alone autotuner integration.


\section{Understand the bit behavior under GPU parallelism}
\label{sec:theory}

\subsection{Dependence Tree of reduction}
\label{sec:theory-reduction}

\begin{figure*}[t]
  \centering
  \includegraphics[width=\textwidth]{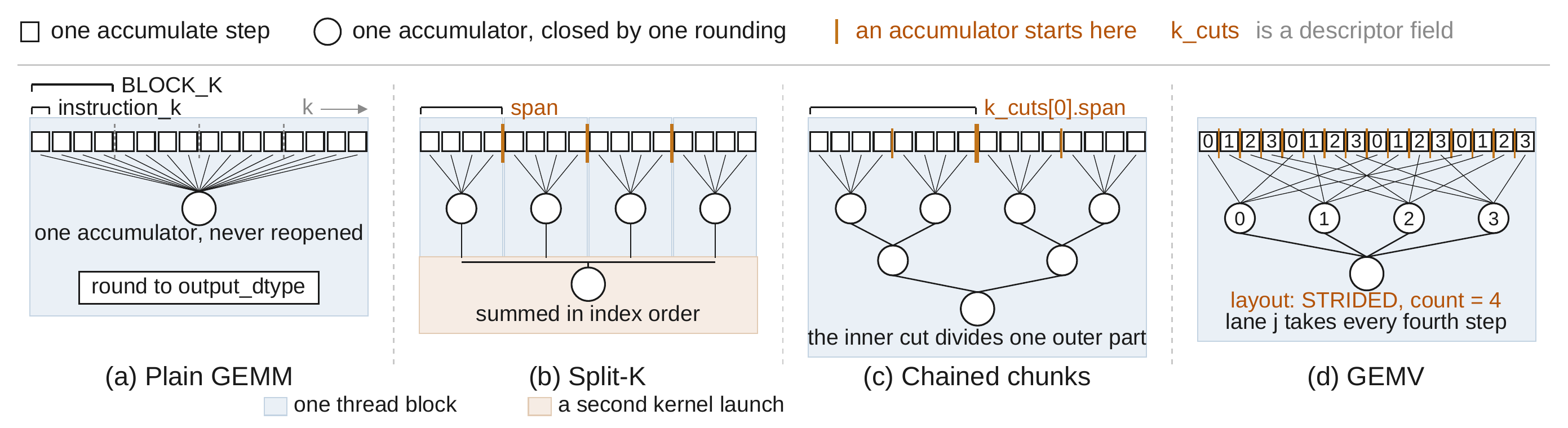}
  \caption{Where each GEMM algorithm family cuts the contracted dimension.}
  \label{fig:gemm-families}
\end{figure*}

An addition of data through a fp32
accumulator it runs is written $\fpadd{fp32}$. 
Addition commutes, but does not associate:
$(a \fpadd{fp32} b) \fpadd{fp32} c \neq a \fpadd{fp32} (b \fpadd{fp32} c)$.
A reduction of $n$ values on a GPU is a directed acyclic graph (DAG). 
Its nodes are of two kinds: a data node holds one floating-point value, 
and an operation node is one floating-point addition. An edge runs from a node to the operation that consumes
its value, so every edge is a data dependence and points the way the value
travels. The graph is a tree in most cases for one reduction result 
(or can be expanded into a tree), 
and we call it the \emph{dependence tree} of the reduction.
The \emph{reduction order} therefore is represented by a dependence tree:
Exchanging an operation's two incoming
edges result a different tree holding the same relation, 
and we call them \emph{equivalent trees}. 
A kernel may induces more than one result, 
and we write $\mathcal{T}(K)$ for the family of their dependence tree. 
Two kernels
over the same inputs return the same bits when the trees at every output
coordinate are equivalent.

A dependence tree also says how much of the reduction can run at once. Its work,
the count of operation nodes, is $n-1$ whatever shape it takes, so a tree's
parallelism, its work divided by its longest dependence
chain, is settled by that chain alone.
Across the trees a compiler may build over the same values the chain runs from
$n-1$, a left fold whose parallelism is 1, down to $\lceil \log_2 n \rceil$ for a
balanced tree. The order
Section~\ref{sec:enforcement} pins is the balanced one, so \textbf{enforcing it gives up
no parallelism chance in theory} ~\cite{blelloch_maggs_parallel, clrs, kuck_muraoka1974}
and still preserves data locality.

A GPU builds that tree in three levels. A thread folds the nodes held in
its own registers, in index order. The threads of a warp fold their partials
through lane exchanges. The warps holding a partial fold theirs through shared
memory. The tree is settled only once all three are fixed,
$\mathcal{T} = \mathcal{T}_{\mathrm{smem}} \circ \mathcal{T}_{\mathrm{warp}}
\circ \mathcal{T}_{\mathrm{reg}}$.
Thread count, the elements one thread holds and the warp count each
decide which nodes share an accumulator at some level, so each of them moves
the tree: Figure~\ref{fig:inner-tree}(a) folds the same eight values under
two layouts and the two roots differ. 
Figure~\ref{fig:inner-tree}(b) folds sixteen values under two layouts, the
second over twice the threads of the first, and the two roots agree.
Section~\ref{sec:enforcement} is how it is pinned by compiler backend.

\textbf{Pipeline depth leaves the tree alone.} A software-pipelined loop
issues the loads for later iterations while an earlier one computes, and the
pipeline depth, \texttt{num\allowbreak \_stages} in Triton, is how many iterations are in
flight at once. What it sets is arrival time. Iteration $i$ still accumulates
iteration $i$'s values into the same accumulator and in the same order, so every
node reaches the accumulator it reached before and the tree is what it was. The
premise is that pipelining moves the loads and leaves the loop's accumulation
chain where it stands.

\textbf{Warp specialization leaves it alone.} A warp-specialized
kernel splits the block's warps into a producer set $P$, which issues
asynchronous copies into shared memory, and a consumer set $C$, which does the
arithmetic. Three facts settle the question. A node of the tree is a
floating-point addition and a producer warp executes none, so $P$ contributes no
node. $\mathcal{T}_{\mathrm{reg}}$ is fixed by the map from a node to the lane
holding it, and that map is computed from the tile shape and the layout, which
the split leaves untouched. Specialization adds the producers alongside the
consumers rather than taking consumers away, and $\mathcal{T}_{\mathrm{smem}}$ spans
the same partials. All three levels are unchanged, so $\mathcal{T}$ is, and so
are the bits. The third fact is a premise about the compiler rather than about
the hardware: a compiler that re-partitioned a tile over fewer compute warps
when it specialized them would break the claim.

\subsection{GEMM reduction}
\label{sec:theory-gemm}

\subsubsection{Tensor core semantics}
\label{sec:theory-tensorcore}

A matrix instruction takes a tile of $A$ and a tile of $B$, multiplies them, and
adds the products into an accumulator that already holds a value. Where the
roundings fall inside that is a black box. For fp16, bf16 and tf32 inputs this
section assumes the semantics below, and every measurement in
Section~\ref{sec:eval} agrees with it: one instruction folds \texttt{instruction\_k} products and the incoming accumulator
into a single rounding:
$\mathrm{acc} \leftarrow \mathrm{acc} \big( \sum^{\fpadd{fp32}}_{j <
\mathtt{instruction\_k}} a_j b_j \big)$,
where the sum is exact, so the instruction's one rounding is the
$\fpadd{fp32}$ itself. The products inside one instruction therefore have
no order of their own, and what survives as a reduction order is the chain of
$\fpadd{fp32}$ steps across instructions, one step per instruction. The scalar
path rounds once per element, $\mathrm{acc} \leftarrow \mathrm{fma}(a,b,\mathrm{acc})$,
and twice when the multiply is left uncontracted. Three instructions carry the
matrix path on NVIDIA: \texttt{mma.sync} in a warp, \texttt{wgmma} in a Hopper
warpgroup, and \texttt{tcgen05.mma} from a single Blackwell thread with its
accumulator in tensor memory. On GB300 \texttt{mma.sync} and
\texttt{tcgen05.mma} are bitwise equivalent at fp16 and at bf16,\footnote{At fp8 the two are not bitwise equivalent, because fp8 accumulation runs at reduced precision with a promotion cadence that need not match across two lowerings~\cite{mmasim, tcmodels}.} so the descriptor names the instructions one by
one rather than carrying a single tensor-core value.

\subsubsection{The algorithm families}
\label{sec:theory-families}

What separates one GEMM from another is where
the contracted axis is cut and what happens at each cut. 
Four GEMM algorithm cover every
library kernel that is used, and Figure~\ref{fig:gemm-families} draws them.

\begin{sloppypar}
\textit{Plain GEMM} hands one output tile to one threadblock, gives every
element of that tile an accumulator, and walks the whole contracted axis inside
that block, the uncut axis of Figure~\ref{fig:gemm-families}(a). Three things about that walk decide the order. A matrix
instruction does not add one product at a time: it folds a fixed number of them
into a single rounding, and that number sets how long the chain of roundings is,
so we record it as \texttt{instruction\_k}. The instruction's own result then has
to reach the accumulator, and it can do so inside that same rounding or in a
second one after it, which is one rounding per step against two; we record which
as \texttt{use\_fast\_accum}. And $K$ is rarely a multiple of the mainloop's
step, so one turn of the loop is short, and whether that short turn runs first
or last changes which products share the first rounding and therefore every
partial sum after it. We record it as \texttt{k\_loop\_step}.

What tiles the output stays out of all three. \texttt{BLOCK\_M} and
\texttt{BLOCK\_N} decide which thread owns which output element, and every
element has an accumulator of its own either way, so re-tiling moves work
between threads and leaves each accumulator's chain where it was.
\texttt{BLOCK\_K} sets how many instructions one turn of the mainloop issues,
and the accumulator is carried across turns, so by the semantics of
Section~\ref{sec:theory-tensorcore} the same instruction results reach it in the
same order however the turns are cut. The three block sizes are speed knobs,
which is why \texttt{instruction\_k} and not \texttt{BLOCK\_K} is the number the
order depends on.

\textit{Split-K}~\cite{agarwal1995} cuts the axis into contiguous parts and gives
each part to a different threadblock, which starts it at zero in an accumulator
of its own, the four parts and their merge in Figure~\ref{fig:gemm-families}(b). We record the length of a part as \texttt{span}. Some kernels cut
again inside a part, closing an accumulator every so many elements before the
part is finished, so what records the cutting is a nest, outermost first, which
we write as \texttt{k\_cuts}. A second
kernel sums the parts.
A finished part is written at one
type and the running sum kept at another, and the two are set independently:
parts at fp32 summed at fp32, parts at the output type summed at fp32, or both
at the output type. We record them as \texttt{partial\_dtype} and
\texttt{merge\_dtype}.

\textit{Chained chunks} cuts twice and stays inside one threadblock, the two
levels of Figure~\ref{fig:gemm-families}(c). Several
threads share one output element, and the outer cut gives each of them a
contiguous chunk of the axis; the inner cut divides a chunk into the sub-blocks
one accumulator closes over. Both lengths are the \texttt{span} of their level of
the nest. This is what a kernel does on the scalar path, where no matrix
instruction folds a group of products for it and the accumulation is an explicit
chain of fused multiply-adds.

\textit{GEMV}~\cite{blas2_1988} multiplies a matrix by a vector. One output
element per row leaves no output tile to spread over threads, so the fold itself
is spread instead, and that makes its order the most exposed of the four. The
lanes of one warp each take every \texttt{count}th tile of the axis rather than a
contiguous slice of it, the deal Figure~\ref{fig:gemm-families}(d) draws, and which of the two a cut does is recorded as
\texttt{layout}. The lanes then fold their totals into one through shuffles, and
that fold is where the order stops being a chain: a butterfly pairs lanes a
power of two apart, which is a balanced tree, and a butterfly that counts its
offsets down rather than up is a balanced tree over the lanes in bit-reversed
order. Those are different sums, and only the second is what these kernels do, so
we record the fold as one two-part cut per round rather than as a single tree
over the lanes.

 We formalise that whole set of factors as the \textbf{GEMM computation descriptor},
\texttt{GEMMDesc} for short, and Section~\ref{sec:reconstruction} is where it is
put to work.
\end{sloppypar}

\subsection{Attention}
\label{sec:theory-attention}

Readers may refer to Appendix~\ref{app:attention} for the reduction order of a
complex kernel such as flash attention, where this work offers a theoretical
description and leaves the implementation open.

\section{Black-box reconstruction}
\label{sec:reconstruction}

Although cuBLAS\footnote{CUDA 13.3 is the latest release at the completion of
this work.} exposes a cost model API that names the GEMM algorithm it would run
for a shape, that answer stops short of the parameters of
Section~\ref{sec:theory-families} that fix the bitwise semantics.

A \texttt{GEMMDesc} is the frozen record holding every one of those factors and
nothing else, so two equal records owe
each other the same bits. What we recover is not one of them. For a fixed
library generation, cuBLAS 12 or cuBLAS 13, we recover the \emph{function}
that generation computes, $(\text{SM version},\, \text{tensor
shape})\allowbreak \longmapsto \allowbreak \texttt{GEMMDesc}$, and a single
shape's descriptor is one point of it. Building that function is
this section's subject: how one point is recovered, and how the points become a
function.

\textbf{Two instruments.} \textbf{\emph{Runtime profiling}} says what the library actually
did: how many kernels it launched, over what grid, and whether it took a
workspace. That narrows a shape to a family cheaply. 
The arithmetic is settled by \textbf{\emph{numerical experiment}}: two extreme
values of opposite sign and one extreme small value, $+L$, $-L$ and $r$, are
placed on an otherwise empty axis. For fp16, $L = 1024$ and $r = 2^{-15}$, which
is under half an ulp of fp32 at $L$. The pair cancels exactly, so $r$ reaches
the output from an accumulator that has already cancelled it and is swallowed by
one still holding half of it. 
A tensor core folds its own $k$ before the accumulator sees anything, so $r$
goes one instruction group away from its $L$ rather than beside it. 
Two placements read the two shapes a grouping
takes. 

\textit{i) Contiguous grouping.} Eight terms summed as
$(a_0 + a_1 + a_2 + a_3) + (a_4 + a_5 + a_6 + a_7)$ answer differently from the
same eight in one chain, and what has to be found is where a group ends. Put $r$
at index $0$ and walk $+L$ and $-L$ as an adjacent pair, at $(1,2)$, then
$(2,3)$, and on. While the pair sits inside $r$'s own group the $+L$ arrives
after $r$ and absorbs it, and the output is zero. At $(3,4)$ the $+L$ is still in
that group and absorbs $r$ there, so the output is zero once more. At $(4,5)$
the pair lies wholly in the second group, which cancels to zero on its own, and
$r$ reaches the output intact. The first placement that returns $r$ is $4$, so
the first group is $a_0$ through $a_3$; pinning $r$ at $4$ and walking again
gives the next boundary.

\textit{ii) Strided grouping.} The same eight summed as
$(a_0 + a_2 + a_4 + a_6) + (a_1 + a_3 + a_5 + a_7)$ hold every other term in one
group, and what has to be found is which terms a group holds. Here the pair is
pinned and $r$ walks. Under the layout being tested, $+L$ goes at a group's
first term and $-L$ at its last, $0$ and $6$ here, so that group carries an
uncancelled $L$ throughout, and $r$ takes each remaining index in turn. It is
absorbed at $2$ and $4$ and survives at $1$, $3$, $5$ and $7$, so the group
holds $\{0, 2, 4, 6\}$: stride two, two groups. A layout guessed wrong puts the
$-L$ outside the group its $+L$ is in, and the pattern that comes back is no
longer periodic.

\textbf{Taking split-K as the example.} Three things that fix split-K's bits
are missing from that answer: \texttt{k\_cuts}, how the axis is cut,
\texttt{span}, how long a part is at each level of that nest, and the pair
\texttt{partial\_dtype} and \texttt{merge\_dtype}. Reading \texttt{span} takes one
row of an fp16 GEMM on GB300. An otherwise empty row of a $K = 576$ GEMM carries
$r$ at $k = 0$ and the pair $+L$, $-L$ at $k = m$ and $m + 1$, walked along the
axis. While the pair shares $r$'s part the $+L$ arrives after $r$ and absorbs
it, so the output is zero from $m = 1$ to $191$. From $192$ the pair lies wholly
in a later part, which cancels to zero on its own, and $r$ reaches the output.
It falls back to zero at $m = 383$ alone, where the pair straddles the next
boundary, so its halves reach the merge apart and the $+L$ absorbs $r$ there.
One walk gives both boundaries, and the axis is in three parts of $192$.

Recovering one point runs the two instruments in order. A profile of the shape
narrows it to a family and drops the candidates that disagree with what was
launched. Each grouping still open is then read off the axis with $+L$, $-L$ and
$r$. What is left over is separated by running each candidate's own walk against
the library on fresh draws until one survives.

\textbf{From points to a function.} A point at a time would never finish:
the shapes are unbounded and the cost model is free to answer differently at each
of them. What makes the function finite is that the descriptor depends on the
shape only through the cost model's answer for it, and those answers are finite
in number. Recovering the function is therefore
recovering one descriptor per reachable answer, which is a table, and we call
one such table an \emph{arch profile}. Following this principle we reconstructed
the tables for fp16, bf16 and fp8 e4m3 on GB300, GB200 and H100 
against cuBLAS 12 and 13, one per SM
version and library generation. The
answers a table has to key on are
enumerated rather than sampled: a scan of 8.25 million cost-model queries over
vector lengths to $10^6$ found every answer the vector family is reached with,
seven of them only at very long vectors or very deep $K$. 
On GB300, GB200, H100, excluding a located cuBLAS bug,
our reconstruction using Triton (block-level programming)
returns cuBLAS's own bytes on every shape whose kernel computes the full sum,
and holds at 100\%.  

A non-cuBLAS GEMM taken alone falls short of highly optimized cuBLAS whether or not 
we require bitwise equivalent:
above 5 GFLOP of arithmetic
the Triton GEMM \tcompile{} generates 
(FP order free) runs at 61--88\% of it over the layer groups, 
while our bit-exact Triton kernel runs at 56--93\%, 
and a large-tensor acceleration (Appendix~\ref{app:accel}) at 69--91\%.
Below 5 GFLOP the free one runs at 59--88\% while ours runs at 50--100\%,
which is where holding the order costs.
\textbf{Kernel fusion} turns that around on realistic shapes: 
over 96 of them the fused pair \tcompile{} generates
runs at 85--125\% of the speed of a cuBLAS GEMM followed by its epilogue as a
second kernel over the epilogue groups, while our bit-exact fused pair runs at
95--168\%. 
Fixing the reduction order is usually expected to cost about 20\% of
matrix-multiply throughput~\cite{he2025defeating}, but these results open the
opposite prospect, that bitwise consistency and performance can be pursued
together.
Section~\ref{sec:eval} carries the complete evaluation result.

\section{Compiler enforcement and layout optimization}
\label{sec:enforcement}

Triton has no balanced tree reduction. Its lowering folds the values a thread
holds one at a time, left to right, and the layout decides which values those
are, so the tree a kernel computes moves with every layout the autotuner tries.
We extend Triton's compiler with one, and with a data layout optimization pass over it,
at an acceptable performance cost against the unordered mode under autotuning.

\subsection{Backend lowering}
\label{sec:enforcement-lowering}

The lane exchange is where the tree is built out of shuffles, and
Figure~\ref{fig:innertree} is the function that emits them. Both modes walk the
same sequence of shuffle-xor steps and differ in direction alone: the pinned
mode counts the offset up from one, so neighbouring lanes pair first and the
tree has the same shape whatever the warp count, while the default counts down
from half the lane count. A switch of this shape sits at every level of
Section~\ref{sec:theory-reduction}'s hierarchy, so the within-thread fold and
the cross-warp stage are pinned the same way.

\begin{figure}[!ht]
\begin{lstlisting}[style=cpp]
void warpReduce(ConversionPatternRewriter &rw, Location loc, 
  SmallVector<Value> &acc, ReduceOp op, unsigned nLane, ...) {
  if (targetInfo.warpReduce(rw, loc, acc, op, nLane, ...)) return;
\end{lstlisting}
\begin{lstlisting}[style=cpp,backgroundcolor=\color{treebg}]
  if (isInnerTree(op)) { // count up: neighbours pair first
    for (unsigned N = 1; N <= nLane / 2; N <<= 1) {
      SmallVector<Value> shfl(acc.size());
      for (unsigned i = 0; i < acc.size(); ++i)
        shfl[i] = targetInfo.shuffleXor(
                    rw, loc, acc[i], N * interleave);
      accumulate(loc, rw, op.getCombineOp(),
                 acc, shfl, pred);
  ...
\end{lstlisting}

\caption{The two orders, in the lane exchange.}
\label{fig:innertree}
\end{figure}

The dispatch above it reaches the pinned path by two routes. The attribute is
one. The other is a layout whose reduced register-lane extent on the axis is
still greater than one, which the default lowering cannot express, so it is sent
down the same path and gets the same tree.

AMD reaches the same tree on different instructions. A wave64 reduction
folds within a row with \texttt{row\_shr} steps, and the two orders are the two
directions through the same step sequence: 8, 4, 2, 1 counting down for the
default, and 1, 2, 4, 8 counting up for the pinned mode. The cross-row broadcast
steps that follow preserve the tree and are identical either way. Where the
assumption behind that sequence fails, on a wave32 target or a partial warp, the
backend declines the instruction-level path and the shared count-up shuffle tree
performs the reduction instead, which is the same tree reached more slowly.

\subsection{The data-layout optimization}
\label{sec:enforcement-layout}

A reduction whose axis is spread across warps pays
a cross-warp stage: two barriers, a shared-memory round trip and a second
shuffle sequence. Moving the operand to a layout that puts the axis inside one warp
escapes that stage, and it normally moves the tree along with it, so the escape
costs bitwise equivalence.

A pinned tree makes the move safe. The lowering above fixes the association
independently of where the elements live, so for a reduction whose tree is
pinned the entire space of valid layouts is a free performance knob.
Algorithm~\ref{alg:layout}, in Appendix~\ref{app:layout}, is the pass that spends it. It builds the layout that
would reduce the axis inside one warp, puts the warps on the kept dimensions so
the axis stays warp-synchronous, carries whatever axis extent is left in
registers as a within-thread fold, and then asks whether that is worth the
\texttt{convert\_layout} it has to insert. The op, its axis, its extent and its
ordering are never touched, so the result is bit-identical by construction.
Figure~\ref{fig:inner-tree}(c) is that rewrite on one operand. Before it a lane
holds a two-by-two block, so the eight-element axis runs past the warp boundary;
after it a lane holds a row of the kept dimension and each warp owns whole
reductions. Loading in the reduce-friendly layout would reach the same place
with no conversion at all, and the pass rejects it: a strided load is paid on
every launch and the conversion once per tile.

On GB300 the optimization recovers most of what the constraint costs, and on
H100 it carries six of ten kernels to or past the unordered mode (see
Section~\ref{sec:eval}).

\section{Static equivalence checking and partition}
\label{sec:checking}

The question the checker answers is narrow: given two compiled kernels, are they
bitwise equivalent. It answers from the assembly statically. 
What it reads is PTX for NVIDIA and AMDGCN for AMD.

\subsection{The algorithm}
\label{sec:checking-algorithm}

\begin{figure}[t]
  \centering
  \includegraphics[width=\columnwidth]{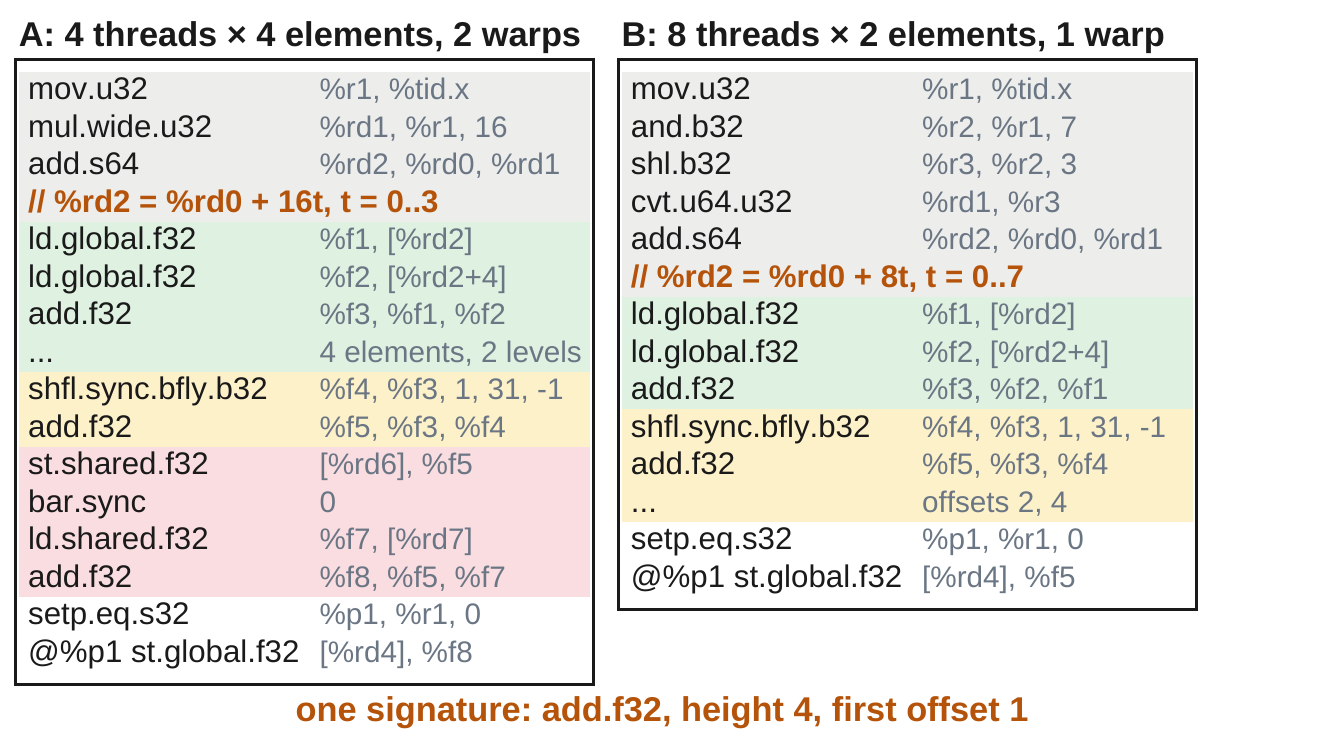}
  \caption{Two layouts of one reduction, one signature.}
  \label{fig:checker-walk}
\end{figure}

The walk takes one entry function in program order over a symbolic thread, one
thread whose index stays a symbol, carrying a map from register to the node
standing for that register's contents. Each instruction is a transfer function on that map:
it looks its operands up rather than tracing them backwards, and writes a node
into its destination. A global load becomes a data node carrying the address the
load read. A floating-point combine becomes an arithmetic node over the nodes
its operand registers already hold. A lane shuffle whose result is combined with
the partial that was shuffled becomes one within-warp exchange node, and a store to
shared memory, followed by a barrier, followed by a load, becomes one cross-warp
exchange node. Figure~\ref{fig:checker-walk} colours two assemblies by scope, thread green,
warp yellow and shared memory pink, as Figure~\ref{fig:inner-tree} colours its
nodes. A loop that
accumulates into a carried register becomes a fold over what one iteration
contributes. A predicate is dropped when its truth is a function of the thread
and block coordinates, since which threads run is a launch fact; a predicate
carrying a loaded value is data, and the entry it guards is compared
configuration by configuration instead. The nodes reaching a global store are
the roots, one per output element, and each carries the dependence tree of that
element.

\textbf{Addresses.} Address registers are evaluated symbolically. 
Over a basis $\mathcal{S}$ of the integer symbols a kernel can read
(the thread and block coordinates, the block extents, the parameters and the
base pointers), every integer register lands in
$\mathcal{A} = \{\, c_0 + \sum_{s \in \mathcal{S}} c_s\, s \,\} \cup \{\, \top_e \,\}$:
an affine form, or an opaque token $\top_e$ carrying the expression that produced
it. Moves and address-space casts are the identity, addition adds termwise, and a
multiply or a left shift by a literal scales every coefficient. A bitwise
operation is admitted only where a rule proves it exact. Writing
$\mathrm{tz}(a)$ for the trailing zeros common to $c_0$ and every $c_s$, a right
shift by $k$ is exact when every term is a multiple of $2^k$, so
$\mathrm{tz}(a) \ge k$ gives $a \gg k = (c_0 \gg k) + \sum_s (c_s \gg k)\, s$;
a mask is the identity when it covers every bit $a$ can set, and an or over two
values that share no set bit is an addition. What is left is a mask or a shift on a thread
index, and the launch bound makes that exact too: $\mathtt{\%ntid}.d = N_d$ is a
power of two, so
$\mathtt{\%tid}.d = \sum_{i < \log_2 N_d} 2^i\, \mathtt{\%tid}.d.\mathrm{bit}_i$
covers exactly $[0, N_d)$, and in that basis a mask keeps the bits it
selects while a shift renumbers them. A leaf's identity is the affine form of
the address it reads, so two kernels reaching one element by different index
arithmetic give one leaf. An opaque token equals only an identical token, which
splits a class rather than merging two.

Canonicalization sorts the children of every commutative node, so a tree is
compared up to exactly the commutativity Section~\ref{sec:theory-reduction}
allows. A bottom-up hash then reduces the tree to one \emph{signature}, and two
kernels are bitwise equivalent when their signatures agree.

A balanced tree reduction collapses one step further. Its shape is fixed by the
logical element order (Section~\ref{sec:theory-reduction}), so the collapsed node
keeps four facts and drops the physical structure that produced them: it keeps
the combine with its rounding, the leaf computation with its coordinate blanked,
the \emph{reduction height}, which is the base-two logarithm of the elements
folded in, and the offset of the butterfly step nearest the leaves; it drops how
many exchanges ran and which thread held which element. A shared-memory exchange
relocates one value and combines nothing, so it adds no height, and the height
then counts elements and holds when the warp count changes.
Figure~\ref{fig:checker-walk} reads the two layouts of
Figure~\ref{fig:inner-tree}(b) out of assembly, sixteen elements over four threads
and two warps and then over eight threads and one: the cut between the scopes
moves, the height of four holds, and the two reach one signature. Balance is the guard, an equal-height pair at
every combine, so a left fold keeps its physical structure and its own class.

A construct the walk does not model becomes \emph{opaque syntax}, matched on
the instruction's own text, which keeps the answer sound (see Appendix).

\subsection{Implementation}
\label{sec:checking-impl}

The NVIDIA checker is about 2,500 lines of Python over a PTX parser: the walk,
the symbolic address evaluator, the tree representation and its canonical form,
and the loop summariser. The AMDGCN checker is its twin over the same tree
representation, with the address evaluator and the exchange recognisers replaced
for that instruction set, and it is graded against a corpus of its own.

Triton's autotuner already takes a pruning predicate, so the PTX checker becomes
one with no change to the autotuner: a search keeps the configurations bitwise
equivalent to a reference, either its own first configuration or one the caller
compiles and hands in, and benchmarks those.

A loop is handled where the walk meets it. A loop-carried accumulation is
recognised by its shape in the assembly, an MMA or a combine whose destination
is also one of its sources, as in \texttt{fma.rn.f32 \%f5, \%f1, \%f2, \%f5}. The walk then replaces the
one-iteration \texttt{add(seed, chunk)} it would otherwise produce with a single
fold node over the chunk, dropping the pre-loop seed, a value every
configuration of one kernel shares. The node is keyed on the loop's own increment, the
\texttt{BLOCK\_N} of a chunked reduction, since re-chunking a scalar fold
regroups the sum. The key comes off for a $K$ loop whose tensor cores accumulate
exactly (Section~\ref{sec:theory-tensorcore}) and whose accumulator nothing
outside the MMAs touches before the loop ends: every \texttt{BLOCK\_K} then
issues the same products into the same accumulator in the same order, so the
chunk size leaves the signature and those configurations merge.

An entry whose reduction the walk could not reconstruct keeps its launch
geometry as the signature, so two kernels that both reconstruct to nothing stay
apart. A reduction whose accumulation crosses a loop back edge is summarised
from one iteration, which is exact for a single fold; a nest of folds would make
that summary a guess, so the loop's own trip constants go into the signature and
two kernels differing only in a split count separate.

Across 51,152 compiled configurations on GB300, 4,500 on gfx942, and
minor-scale benchmarks on GB200 and H100, equivalence the checker certified is
equivalence the hardware confirmed. 
On most kernels, the TorchInductor-generated ones
included, it hands back 1.0 to 2.5 times the classes that really exist
(Section~\ref{sec:eval}).

\section{Evaluation}
\label{sec:eval}


\begin{figure*}[t]
  \centering
  \includegraphics[width=\textwidth]{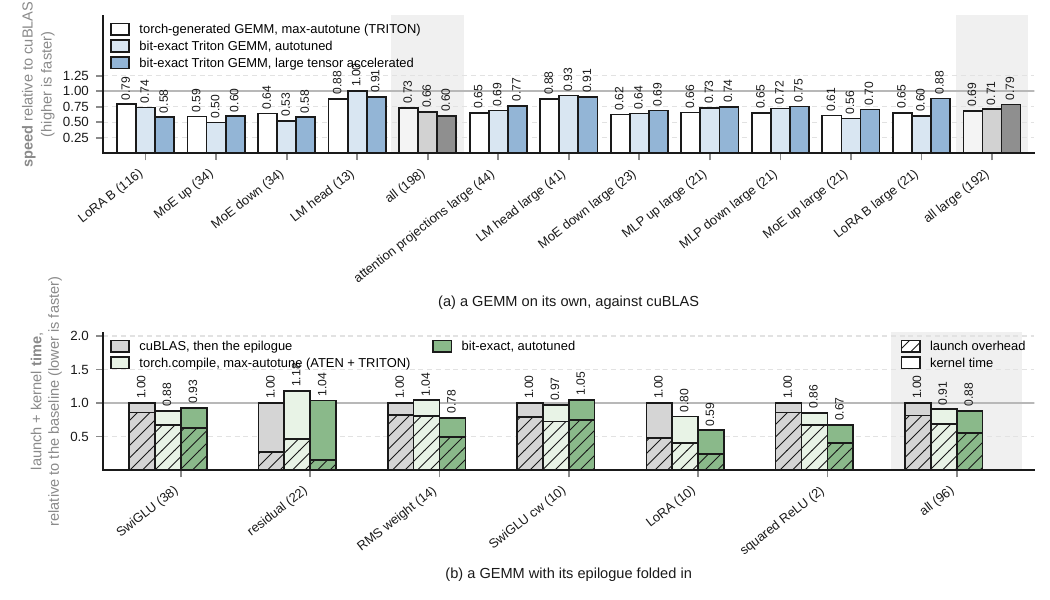}
  \caption{What the bitwise constraint costs, on a GEMM and on a fused epilogue.}
  \label{fig:speed}
\end{figure*}

GB300 (sm\_103) carries every measurement below. The cuBLAS reconstruction and
the checker's soundness gate run on GB200 (sm\_100) and H100 (sm\_90) too, H100
carries the layout optimization as well, and gfx942 (CDNA3) carries the AMDGCN
checker over a corpus of its own. The evaluation splits in two: bit-level
correctness first, then what holding it costs.

\subsection{Benchmarks}
\label{sec:eval-benchmarks}

cuBLAS 13, Triton 3.8, PyTorch 2.12 and CUDA 13 throughout; bit-identical means
the outputs agree byte for byte on every input draw. \textbf{Shapes.} A measurement that feeds a kernel draws its shapes from one of
two sets. The realistic static set is 390 fp16 $(M, N, K)$ triples read from the
layer dimensions of open-weight models, each carrying its layer and the
operation that follows it, with the ranges in Table~\ref{tab:shapes}. The
fuzzing set is drawn at random over two dtypes and reaches the corners a layer
list never visits, $M = 1$ and a $K$ in the hundreds of thousands: 110,813
shapes for the reconstruction and 1,400 for the performance work. \textbf{Kernels.} A measurement that needs the kernel itself takes it from three
sources: micro-kernels written for this work, each isolating one ordering or
layout question; the Triton kernels
\textbf{TorchInductor}~\cite{pytorch2_2024} emits, copied verbatim; and a zoo of
97 kernels adapted from open-source complex
benchmarks~\cite{triton_tutorials,tritonbench,flaggems,flashlinearattention,torchao}. The ordering work runs 24 (kernel, dtype) pairs
over the first two sources at fp16, fp8 and fp32, plus a LayerNorm
weight-gradient reduction from the zoo; the checkers are graded on one PTX file
per autotuner configuration, drawn from all three.

\begin{table}[!ht]
  \centering
  \caption{The static GEMM shapes, by layer.}
  \label{tab:shapes}
  \footnotesize
  \setlength{\tabcolsep}{2pt}
  \begin{tabular}{lrrrr}
    \toprule
    layer                  & shapes & $M$ & $N$ & $K$ \\
    \midrule
    attention projections  &  45 &   256--16,384 &   2,048--18,432 &  2,048--16,384 \\
    MLP up                 &  21 &   256--16,384 &   6,144--33,792 &   2,048--7,168 \\
    MLP down               &  21 &   256--16,384 &    2,048--7,168 &  6,144--33,792 \\
    MoE up                 &  55 &     16--3,072 &      512--3,072 &   2,048--8,192 \\
    MoE down               &  57 &     16--6,144 &    2,048--8,192 &     512--3,072 \\
    LM head                &  54 &        1--256 & 100,352--248,320 &  2,048--8,192 \\
    LoRA B                 & 137 &    80--16,384 &   768--32,768 &          8--64 \\
    \midrule
    all                    & 390 &     1--16,384 &  512--248,320 &        8--33,792 \\
    \bottomrule
  \end{tabular}

  \vspace{2pt}
  \parbox{\columnwidth}{\footnotesize Read from the layer dimensions of 16 open-weight
  models: Qwen3.8-2.4T-A95B, Qwen3.8-27B, Qwen3.6-35B-A3B, Qwen3-30B-A3B, DeepSeek-V4
  Pro and Flash, Kimi-K2.6 and K3, GLM-5.2, GLM-4.7-Flash, gpt-oss-120b and 20b,
  Ling-3.0-flash, MiniMax-M2, Nemotron-3.5-L, and granite-4.1-8b. Every dimension is a
  named key in that model's own \texttt{config.json};
  Appendix~\ref{app:models} gives them per model.}
\end{table}

\subsection{Evaluation on bit-level correctness \& static checker soundness}
\label{sec:eval-correctness}

\textbf{Reconstructing cuBLAS.} Table~\ref{tab:bitmatch} puts a Triton GEMM
against cuBLAS on 110,813 random fp16 shapes on GB300, at ten input draws each.
Every shape whose cuBLAS kernel computes the full sum comes back byte for byte
identical, across all eight algorithm families the library dispatches to. The
60 shapes in the last column are one defect in cuBLAS itself: at very deep contraction lengths the
library sums a whole number of blocks and drops the tail, which we reproduce
with no Triton in the picture. The same holds at fp8 and on the two
older generations: 100\% bit-identical on every shape outside that defect.

\begin{table}[!ht]
  \centering
  \caption{Reconstructing cuBLAS bit for bit, on three GPU generations.}
  \label{tab:bitmatch}
  \footnotesize
  \setlength{\tabcolsep}{2pt}
  \begin{tabular}{lrrr}
    \toprule
    GEMM algorithm (cuBLAS kernel) & tested & bit identical & cuBLAS bug \\
    \midrule
    Single-pass accumulation (\texttt{nvjet})   &  19,449 &  19,449 &  0 \\
    Split-K (\texttt{nvjet})                    &  18,636 &  18,576 & 60 \\
    Per-MMA accumulation (\texttt{cutlass})     &  13,473 &  13,473 &  0 \\
    Split-K, per-MMA (\texttt{cutlass})         &  33,677 &  33,677 &  0 \\
    Three-level chain (\texttt{gemmSN\_NN})     &   4,731 &   4,731 &  0 \\
    Lane-tree GEMV (\texttt{gemv2T})            &  14,897 &  14,897 &  0 \\
    Contiguous-slice GEMV (\texttt{gemv2T})     &   1,722 &   1,722 &  0 \\
    Workspace GEMV (\texttt{reduce\_1Block})    &   4,228 &   4,228 &  0 \\
    \midrule
    total                                       & 110,813 & 110,753 & 60 \\
    \bottomrule
  \end{tabular}

  \vspace{2pt}
  \parbox{\columnwidth}{\footnotesize GB300 (sm\_103) against cuBLAS 13, fp16, 110,813 random
  shapes at ten input draws each, from six regimes spanning $M$ and $N$ from 1
  to 120,000 and $K$ from 8 to 300,000. Appendix~\ref{app:cublasbug} is the
  defect behind the last column, with its standalone reproduction.}

  \vspace{6pt}
  \begin{tabular}{lrrr}
    \toprule
    architecture & tested & bit identical & cuBLAS bug \\
    \midrule
    GB300, fp8        &  42,793 & 99.93\% & 0.07\% \\
    GB200, fp16 + fp8 & 626,522 & 99.81\% & 0.19\% \\
    H100, fp16 + fp8  & 648,720 & 99.82\% & 0.18\% \\
    \bottomrule
  \end{tabular}

  \vspace{2pt}
  \parbox{\columnwidth}{\footnotesize The fp8 half of the same GB300 campaign, then the same
  evaluation on GB200 (sm\_100) and H100 (sm\_90), whose records hold the two
  dtypes together. Counts are shapes; rates are over byte comparisons.}
\end{table}

\textbf{Soundness.} Over 51,152 configurations on GB300, from 47 Triton kernels in
five floating-point formats, every pair the checker certified as bitwise
equivalent returned the same bytes on every input draw. The same gate holds over the AMDGCN
checker's own 4,500 configurations, and over minor-scale benchmarks on GB200 and
H100. Table~\ref{tab:checker}
reads that corpus one dtype per kernel: on the reductions and GEMMs this
paper targets the checker over-splits the true partition by 1.0 to 2.5, reaching
it exactly on the GEMM family, and four kernels outside of our major focus
run from 6.7 to 23.0.

\begin{table}[!ht]
  \centering
  \caption{Static equivalence checking on NVIDIA PTX and on AMD GCN.}
  \label{tab:checker}
  \footnotesize
  \setlength{\tabcolsep}{1.2pt}
  {\fontsize{6.2}{7.2}\selectfont
  \begin{tabular}{lrrrrr}
    \toprule
                 &         & \multicolumn{2}{c}{classes} & over-  & over-  \\
    \cmidrule(lr){3-4}
    kernel       & configs & checker & bytes & split  & merges \\
    \midrule
    reductions on different dims/axes (16), fp16
      & 2,304 \amd{1,920} & 204 \amd{144} & 81 \amd{87} & 2.5 \amd{1.7} & 0 \amd{0} \\
    \texttt{softmax}, fp16
      & 144 \amd{120} & 8 \amd{10} & 4 \amd{3} & 2.0 \amd{3.3} & 0 \amd{0} \\
    \texttt{layernorm}, fp16
      & 144 \amd{120} & 19 \amd{20} & 9 \amd{3} & 2.1 \amd{6.7} & 0 \amd{0} \\
    \texttt{rmsnorm}, fp16
      & 144 \amd{120} & 17 \amd{20} & 8 \amd{3} & 2.1 \amd{6.7} & 0 \amd{0} \\
    \texttt{gemm}, fp16
      & 6,304 \amd{72} & 1 \amd{4} & 1 \amd{4} & \textbf{1.0} \amd{\textbf{1.0}} & 0 \amd{0} \\
    \texttt{gemm\_bias\_relu\_fp\_fusion}, bf16
      & 192 \amd{24} & 1 \amd{2} & 1 \amd{2} & \textbf{1.0} \amd{\textbf{1.0}} & 0 \amd{0} \\
    \texttt{gemm\_kgroup}, bf16
      & 1,152 \amd{132} & 138 \amd{1} & 6 \amd{1} & 23.0 \amd{\textbf{1.0}} & 0 \amd{0} \\
    \texttt{gemm\_reduce\_sum}, fp32
      & 270 \amd{--} & 60 \amd{--} & 9 \amd{--} & 6.7 \amd{--} & 0 \amd{--} \\
    \texttt{gemm\_softmax}, fp32
      & 270 \amd{--} & 60 \amd{--} & 9 \amd{--} & 6.7 \amd{--} & 0 \amd{--} \\
    \texttt{gemm\_tma\_store}, fp16
      & 18 \amd{12} & 1 \amd{1} & 1 \amd{1} & \textbf{1.0} \amd{\textbf{1.0}} & 0 \amd{0} \\
    \texttt{inductor\_sum\_loop}, fp16
      & 20 \amd{--} & 20 \amd{--} & 20 \amd{--} & \textbf{1.0} \amd{--} & 0 \amd{--} \\
    \texttt{inductor\_splitk\_gemm}, fp16
      & 4 \amd{--} & 4 \amd{--} & 4 \amd{--} & \textbf{1.0} \amd{--} & 0 \amd{--} \\
    \texttt{flash\_attention}, fp16
      & 2,720 \amd{--} & 2,655 \amd{--} & 266 \amd{--} & 10.0 \amd{--} & 0 \amd{--} \\
    \bottomrule
  \end{tabular}}

  \vspace{2pt}
  \parbox{\columnwidth}{\footnotesize Black is the PTX checker on GB300;
  \textcolor{amdblue}{blue, in brackets, is the AMDGCN checker on gfx942} over a
  corpus of its own, where a dash is a kernel that corpus does not carry.}
\end{table}

\subsection{Evaluation on performance cost}
\label{sec:eval-cost}

\textbf{A GEMM on its own.} Panel (a) of Figure~\ref{fig:speed} splits the
390 shapes at 5 GFLOP of arithmetic ($2MNK$), where a GEMM gains enough work to
saturate the machine. Above it, where the Triton GEMM
\tcompile{} generates with the order left free reaches 61--88\% of
cuBLAS over the layer groups,\footnote{\tcompile{} emits a Triton GEMM
of its own rather than reproducing cuBLAS's algorithm, so it is not necessarily
the faster kernel.} the bit-exact GEMM reaches 56--93\% and the
large-tensor accelerated one 69--91\%. Below the line the free one reaches
59--88\% and ours 50--100\%.

\textbf{A GEMM with its epilogue.} Panel (b) folds the operation that follows
each GEMM in its model's own code into the kernel, over 96 shapes in six
epilogue groups. Each bar is a call's launch overhead plus its kernel time over
the same two parts of a cuBLAS GEMM and a separate epilogue kernel, where launch
is 82\% of the baseline. Where \tcompile{} reaches 85--125\% of that
baseline, the bit-exact fused kernel reaches 95--168\%: it runs 1.8 times as long
as the baseline's kernel half and still wins the pair by removing one launch.

\begin{figure}[!ht]
  \centering
  \includegraphics[width=\columnwidth]{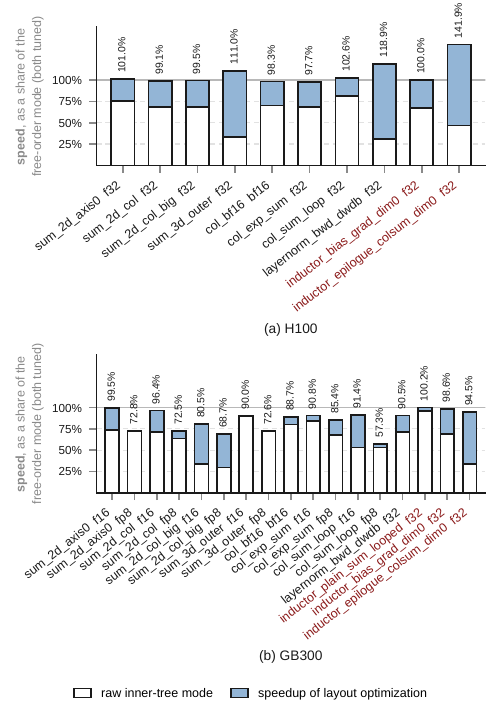}
  \caption{What the reduction-ordering constraint costs, and what the layout
  optimization takes back.}
  \label{fig:layout}
\end{figure}

\textbf{The enforced reduction order.} Figure~\ref{fig:layout} reads the fix-bits layout
optimization against the free-order mode, which lets the compiler pick any
reduction order. Both modes are tuned, so each bar is that mode's own best; 
the white part of a bar is where the ordering constraint alone
leaves the kernel. 19 of the 27 bars finish within 10\% of the
free-order mode, and on H100 the optimization carries six of the ten kernels to
or past it, by as much as 42\% on the column sum a matrix-multiply epilogue
emits. On GB300 it passes that mode once and otherwise reaches 57\% to 99\% of
it. 

\section{Discussion and future work}
\label{sec:discussion}

Several limitations remain, and each is a future direction this work leaves open.
\emph{i)} Section~\ref{sec:theory} says enough about bit-level semantics for a
descriptor to hold across machines and across library versions: none of its
fields names a machine, and the split is written down rather than read off the
hardware, so it survives the change in SM count at which cuBLAS's own guarantee
lapses~\cite{cublas_docs}. Measuring that cross-machine robustness 
is what the resources behind this work
did not reach, due to resource limitations.
\emph{ii)} Taming a GEMM's arithmetic could go up a level, from one kernel to a
system that composes many of them; and how much contribution this work can benefits the 
ML/RL training job is not measured either due to resource limitation.
\emph{iii)} It could go down a level equally, from the PTX and AMDGCN the
checkers read to the machine code \texttt{ptxas} emits below them: every
soundness claim here is relative to the assembly, and a checker reading SASS
would close that gap at the cost of a format the vendor does not document.
\emph{iv)} Attention and the GEMM fusions whose epilogue impact the
reduction order does not deeply covered but only a theoretical analysis in
Appendix~\ref{app:attention};
\emph{v)} The principle could be built into automatic kernel generation: a
generator such as TorchInductor~\cite{pytorch2_2024} could pin the order it
emits and tune inside one bitwise-equivalence class.

\section{Conclusion}
\label{sec:conclusion}

A GPU kernel's bit-level semantics are decided by the structure of its
floating-point accumulation, and that structure is chosen by a compiler and a hardware mapping
rather than by whoever wrote the kernel. This paper makes the structure a thing
that can be handled. It can be written down, as a descriptor that holds what
decides the bit-level semantics. 
It can be recovered from a closed library by
profiling and numerical experiment, our first black-box reconstruction to reach
bit-level correctness, which reproduces cuBLAS's own bytes across three GPU
generations. It can be requirement of a compiler, and we show that such
requirement costs can be won back at acceptable rate. 
And it can be decided statically from compiled code, soundly, by our first
checkers to settle bitwise equivalence between compiled GPU kernels on two
vendors' instruction sets, over tens of thousands of autotuner configurations,
and that decision goes back to the autotuner, which then searches inside one
class. 
Our work was also heavily evaluated at the view of performance cost, 
which is much more acceptable 
than what the community usually worried on enforcing numerical correcteness.

\appendix


\section{High level graph}
\label{app:overview}

\begin{figure*}[t]
  \centering
  \includegraphics[width=\textwidth]{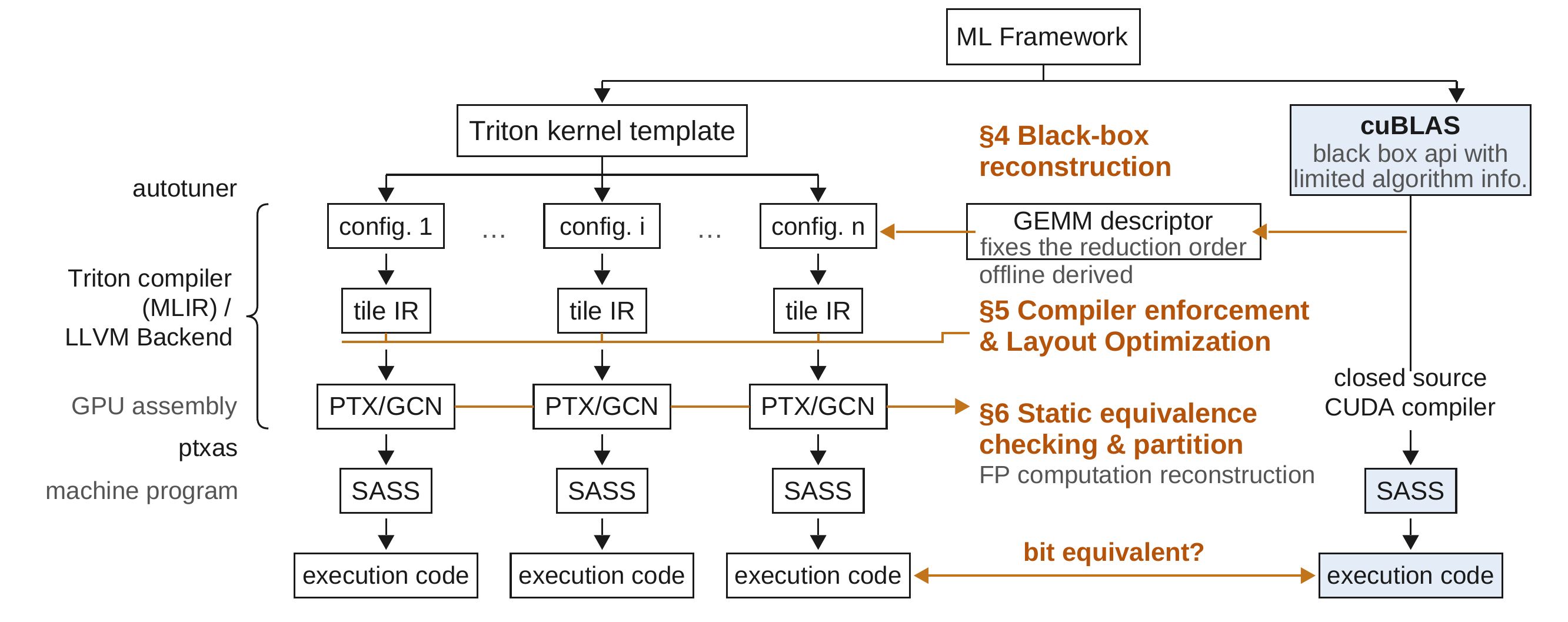}
  \caption{High level overview of this work.}
  \label{fig:big-picture}
\end{figure*}

Figure~\ref{fig:big-picture} places the four mechanisms on the path a kernel
takes from a template to the bytes it returns, beside the closed library whose
arithmetic Section~\ref{sec:reconstruction} reconstructs.

\section{The layout optimization}
\label{app:layout}

The pass takes three numbers besides the module: the warp size $w$, a cap $c$ on
the elements one thread may hold once the operand has moved, and a spread factor
$u$. Every test is a skip, so a reduction it cannot place is left exactly as it
was. A reduction crossing a thread block is out of scope, and one already inside
a warp has nothing to win. The last two guards ask whether the conversion pays:
the candidate has to put more lanes on the axis than the current layout does,
and the axis has to be under-spread by a factor $u$ before the conversion earns
its cost.

\begin{algorithm}[H]
\caption{The layout optimization for a pinned reduction.}
\label{alg:layout}
\small
\begin{algorithmic}[1]
\Require module $M$, warp size $w$, per-thread element cap $c$, spread factor $u$
\ForAll{reductions $r \in M$ with \texttt{reduction\_ordering} $=$ \texttt{inner\_tree}}
  \State $L \gets$ operand layout of $r$; \textbf{skip} unless $L$ is blocked
  \State \textbf{skip} if $r$ crosses a thread block, or is already warp-synchronous
  \State \textbf{skip} if $|{\rm tile}| / (w \cdot n_{\rm warps}) > c$
         \Comment{the relayout would spill}
  \State $C \gets \textsc{Ideal}(r)$: lanes onto the axis first and the rest onto
         the kept dimensions; warps onto the kept dimensions only; the remaining
         axis extent into registers
  \State \textbf{skip} if $C = L$, or if $C$ puts no more lanes on the axis than
         $L$ does
  \State \textbf{skip} if ${\rm extent}({\rm axis}) < u \cdot L.{\rm lanes}({\rm axis})$
         \Comment{already well spread}
  \State convert each operand to $C$, clone $r$ onto it, convert the results back
\EndFor
\end{algorithmic}
\end{algorithm}

\section{The models the shapes are read from}
\label{app:models}

A row of Table~\ref{tab:models} becomes a set of GEMM shapes by reading the
weight each layer multiplies by. In $M \times N \times K$, $N$ is that weight's
output width, $K$ its input width, and $M$ a token count. Attention projections
take their widths from \emph{hidden} and from the head counts in the same file;
the MLP pair takes $N$ from \emph{FFN} going up and $K$ from \emph{FFN} coming
down; the MoE pair does the same with \emph{expert} in place of \emph{FFN}; and
the LM head takes $N$ from \emph{vocab}, which is where the widest shapes in
Table~\ref{tab:shapes} come from.

A LoRA adapter contributes the second of its two matrix multiplications, whose
$K$ is the adapter rank rather than a model width. That is why the LoRA rows of
Table~\ref{tab:shapes} reach down to $K = 8$ while their $N$ stays at
\emph{hidden} or \emph{FFN}, and it is the one layer group whose contracted
dimension a model's configuration does not fix.

\begin{table*}[t]
  \centering
  \caption{The open-weight models the static shapes are read from.}
  \label{tab:models}
  \footnotesize
  \setlength{\tabcolsep}{3pt}
  \begin{tabular}{lrrrrrr@{\hskip 1.2em}lrrrrrr}
    \toprule
    model & hidden & FFN & expert & experts & active & vocab &
    model & hidden & FFN & expert & experts & active & vocab \\
    \midrule
    Qwen3.8-2.4T-A95B & 8192 & --    & 2048 & 512 & 10 & 248,320 &
    gpt-oss-20b       & 2880 & --    & 2880 &  32 &  4 & 201,088 \\
    DeepSeek-V4-Pro   & 7168 & --    & 3072 & 384 &  6 & 129,280 &
    Nemotron-3.5-L    & 2688 & 1856  & 1856 & 128 &  6 & 131,072 \\
    DeepSeek-V4-Flash & 4096 & --    & 2048 & 256 &  6 & 129,280 &
    MiniMax-M2        & 3072 & --    & 1536 & 256 &  8 & 200,064 \\
    GLM-5.2           & 6144 & 12288 & 2048 & 256 &  8 & 154,880 &
    Qwen3-30B-A3B     & 2048 & 6144  &  768 & 128 &  8 & 151,936 \\
    GLM-4.7-Flash     & 2048 & 10240 & 1536 &  64 &  4 & 154,880 &
    Ling-3.0-flash    & 2560 & 6144  &  768 & 512 &  8 & 157,184 \\
    Kimi-K2.6         & 7168 & 18432 & 2048 & 384 &  8 & 163,840 &
    Qwen3.6-35B-A3B   & 2048 & --    &  512 & 256 &  8 & 248,320 \\
    Kimi-K3           & 7168 & 33792 & 3072 & 896 & 16 & 163,840 &
    Qwen3.8-27B       & 5120 & 17408 & --   & --  & -- & 248,320 \\
    gpt-oss-120b      & 2880 & --    & 2880 & 128 &  4 & 201,088 &
    granite-4.1-8b    & 4096 & 12800 & --   & --  & -- & 100,352 \\
    \bottomrule
  \end{tabular}

  \vspace{2pt}
  \parbox{\textwidth}{\footnotesize Every number is the value of a named key in that
  model's own \texttt{config.json}, read from the Hugging Face
  hub~\cite{huggingface_hub} on 2026-08-16. \emph{FFN} is the dense feed-forward width
  and \emph{expert} the routed one; a dash means the key is absent, so a model with no
  \emph{FFN} entry is mixture-of-experts in every layer and one with no \emph{expert}
  entry is dense in every layer. \emph{active} is how many experts a token is routed
  to. The 16 are drawn from the highest trending and most downloaded models in that
  hub's text-generation listing on the day the set was fixed.
  DeepSeek-V4~\cite{deepseekv4} appears in two sizes.}
\end{table*}

\section{The cuBLAS bug on split-K tail}
\label{app:cublasbug}

The defect sits in the \texttt{nvjet} split-K path cuBLASLt reaches at
\texttt{ALGO\_ID} 66. It reproduces on GB200, GB300 and H100, and under cuBLAS
12.8.5 as well as 13.1.1. Which shape loses its tail moves with the architecture
and the library together, so Table~\ref{tab:cublasbug} gives three that separate
the pairings.

\begin{figure}[t]
  \centering
  \includegraphics[width=\columnwidth]{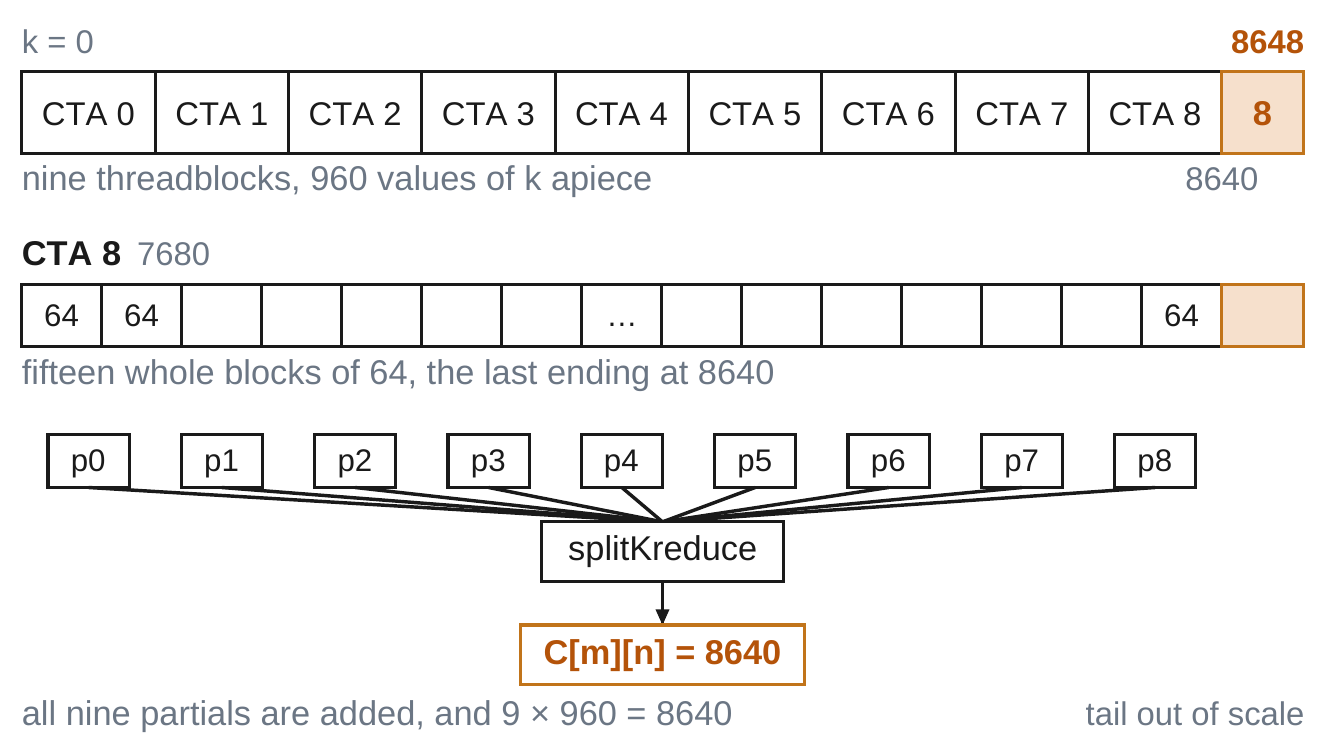}
  \caption{Where the last eight values of $k$ go, at $K = 8648$.}
  \label{fig:cublastail}
\end{figure}

Fill $A$ and $B$ with ones. Every element of $C$ must then be exactly $K$, being
a sum of $K$ products of one with one. A product of two ones is exact in fp32,
the accumulator is fp32, and an fp32 output holds every whole number below
$2^{24}$, so the number that comes back is the count of $k$ values that were
summed. On some shapes it comes back short by a whole number, which is that many
missing terms of $1 \times 1$.

\textbf{Where the piece is lost.} Figure~\ref{fig:cublastail} draws $K = 8648$.
The $k$ step of one threadblock in this fp16 kernel family is 64, and 8648 holds
135 whole steps with 8 values over. cuBLASLt spreads the work over nine
threadblocks, and $135 / 9 = 15$ exactly, so each takes 15 whole steps, or 960
values of $k$. Every threadblock is full, and the last step of the last one is a
whole 64 ending at 8640. The reduction then adds all nine partial results, so
everything computed reaches the output and nothing is added twice. The nine
ranges cover $[0, 8640)$, and $K$ is 8648.

\textbf{The condition.} Write $b$ for the $k$ step, $q = \lfloor K/b \rfloor$,
$t = K \bmod b$, and $s$ for the split count cuBLASLt picks. The tail is lost
exactly when $\mathtt{ALGO\_ID} = 66$, $t \neq 0$, $q \bmod s = 0$ and $s > t$.
At $K = 8648$ that reads $q = 135$, $t = 8$, $s = 9$. The last clause is what
makes $q$ divide evenly into ranges of whole steps with the tail left over: a
split that took an uneven share would give one threadblock the short step and
sum it. The \texttt{ALGO\_ID} clause carries its own weight. On H100 the shape
$1 \times 2 \times 1032$ satisfies all three arithmetic clauses and loses
nothing, because it lands on \texttt{ALGO\_ID} 23, a CUTLASS kernel rather than
an \texttt{nvjet} one. Over 1,542,555 real GEMMs per library on H100 the
four-clause condition predicted every loss with no false positives and no false
negatives, and dropping the \texttt{ALGO\_ID} clause misclassified 3,557 shapes
under each library.

\begin{table}[t]
  \centering
  \caption{Which shape loses its tail on which pairing.}
  \label{tab:cublasbug}
  \footnotesize
  \setlength{\tabcolsep}{4pt}
  \begin{tabular}{lccccc}
    \toprule
        & GB200  & GB300  & GB300  & H100   & H100 \\
    $K$ & 13.1.1 & 13.1.x & 12.8.5 & 13.1.1 & 12.8.5 \\
    \midrule
     8,648 & $-8$ & $-8$ & ok   &      & ok   \\
    57,608 &      & ok   & $-8$ & ok   & $-8$ \\
    11,528 &      &      &      & $-8$ & $-8$ \\
    \bottomrule
  \end{tabular}

  \vspace{2pt}
  \parbox{\columnwidth}{\footnotesize A dash of $-8$ is eight values of $k$ left
  out of the sum, ok is the full sum, and a blank cell was not measured.
  Architectures are sm\_100, sm\_103 and sm\_90.}
\end{table}

\textbf{Which library, which machine.} Table~\ref{tab:cublasbug} gives three
shapes and where each one loses. No single shape covers every pairing, because
the loss needs cuBLASLt to split $K$ at all and that decision moves with both
the architecture and the library. On GB300 the two libraries were nearly
disjoint: over 12,282 shapes, 13.1.1 met the condition on 193 and 12.8.5 on 309,
with no overlap, and at $K = 8648$ the older library returns a split count of one
and stays whole. On H100 they nearly coincide instead, and 11,528 is the
smallest losing $K$ for both. Every pairing runs the same tile family
\texttt{64x8\_64x16\_1x1} in the \texttt{nvjet} split-K path, followed by one
\texttt{splitKreduce} launch.

\textbf{The reproducer.} It calls cuBLASLt directly, since
\texttt{torch.matmul} never reaches this algorithm on its own path. Two things
in it are load bearing. The workspace is where split-K writes its partial
results, so a zero-size workspace keeps cuBLASLt whole and both shapes come
back right. The output is fp32 because at part two's depth an fp16 output
cannot be trusted: above 16,384 the fp16 step is 16, so a $K$ of the form
$64q + 8$ sits
halfway between two fp16 values and rounds down on its own. The control is
$K = 57{,}616$, a shape whose sum is complete, which an fp16 output reads as
57,600 and an fp32 output reads correctly. Asking for fp32 leaves the heuristic's
choice alone, checked over 600 random shapes that returned a bit-identical
configuration either way.

\begin{lstlisting}[style=python]
M, N, K = 1, 8, 8648   # q = 135, t = 8, splits 9
a = torch.ones(M, K, dtype=torch.float16, device="cuda")
b = torch.ones(K, N, dtype=torch.float16, device="cuda")
c = torch.empty(M, N, dtype=torch.float32, device="cuda")

# algo is NULL, so cuBLASLt picks the split; the
# workspace is what lets it split at all
cublasLtMatmul(handle, desc, one, a, a_layout,
               b, b_layout, zero, c, c_layout,
               c, c_layout, None,
               workspace, WORKSPACE_BYTES, stream)

assert c[0, 0].item() == K    # reads back 8640
\end{lstlisting}

\textbf{How far it reaches.} The three shapes are the visible end of a sweep.
Over 496,906 random shapes on a GB300 under 13.1.1, 1,062 came back other than
bit-identical to a full-$K$ reference and 1,059 of those meet the condition;
under 12.8.5 on the same GPU, 179 shapes were flagged and all 179 lose exactly
$t$ values of $k$, while 9,525,600 GEMMs at smaller $K$ stay whole. On H100 the
counts are 2,714 under 13.1.1 and 2,995 under 12.8.5, each out of 1,542,555 real
GEMMs. Every one of the 5,709 observed losses was exactly $K \bmod 64$. It
reaches ordinary shapes as well as skinny ones: 363 of 885 $(M, N)$ pairs under
13.1.1 and 416 of 885 under 12.8.5 lose $k$ somewhere, and $32 \times 32$,
$64 \times 64$ and $128 \times 128$ all lose 8 at $K = 11{,}528$.

\section{The large-tensor acceleration on GB300}
\label{app:accel}

A descriptor fixes the arithmetic and says nothing about the schedule, so a
second kernel that realises one descriptor may choose any schedule it likes and
still return the same bytes. The accelerated arm is that freedom taken up on
GB300: each launcher gained one branch at the top, taken on sm\_103 and falling
through to the original body everywhere else, so the kernels the other two
generations run are the ones they ran before.

The branch carries a \texttt{GROUP\_M} tile swizzle, tensor-memory
descriptors, a persistent grid with warp specialisation, deeper pipelining,
32-bit indexing wherever an operand cannot reach $2^{31}$ elements, a
copy-free operand path, a single launch across the split-K slices, and a tile
rule for each of the eight plan modes a GB300 reaches.

\begin{table}[t]
  \centering
  \caption{What the acceleration is worth, by plan mode.}
  \label{tab:accel}
  \footnotesize
  \begin{tabular}{lrr}
    \toprule
    plan mode & reference & accelerated \\
    \midrule
    single-pass accumulation      & 0.303 & 0.802 \\
    split-K                       & 0.426 & 0.857 \\
    split-K, grouped              & 0.624 & 1.639 \\
    per-MMA accumulation          & 0.654 & 2.086 \\
    three-level chain             & 0.184 & 0.998 \\
    lane-tree GEMV                & 0.210 & 0.843 \\
    contiguous-slice GEMV         & 0.596 & 0.962 \\
    workspace GEMV                & 0.853 & 1.057 \\
    \bottomrule
  \end{tabular}

  \vspace{2pt}
  \parbox{\columnwidth}{\footnotesize Device geometric mean of cuBLAS's kernel
  time over the arm's, by CUDA-graph replay with L2 flushed between replays, on
  an idle GB300. Above 1.0 is faster than cuBLAS, and every such entry is in the
  unaligned regime, where the win comes from repacking the operands to restore
  alignment.}
\end{table}

One change needed an argument that no value moves. The per-MMA plan accumulates
in groups of \texttt{KPD} real $k$ elements, and the reference kernel spends one
\texttt{tl.dot} on each group. The accelerated kernel spends one on $G$ groups
at once. A \texttt{tl.dot} whose $k$ extent is $16G$ lowers to $G$ chained
$k=16$ MMAs on one fp32 accumulator in increasing $k$, and each rounds the
accumulator exactly once, so lanes $[16g,\,16g+16)$ of the wide dot are the
accumulator update the $g$-th narrow dot performed. Where a group is narrower
than an MMA, at \texttt{KPD} 8, the tail of each group is masked to zero, which
is the stand-in the narrow kernel already used, repeated $G$ times in one tile.
The residue tile keeps one group per dot, since its last group can be short:

\begin{lstlisting}[style=python]
# residue tile: one group per dot, the last one may be short
for g in tl.range(0, nfirst, num_stages=1):
    off  = g * KPD
    real = tl.arange(0, 16) < tl.minimum(KPD, rbk - off)
    kk   = k0 + off + tl.arange(0, 16)
    a = tl.load(A + om[:, None]*am + kk[None, :]*ak,
                mask=real[None, :], other=0.0)
    b = tl.load(B + kk[:, None]*bk + on[None, :]*bn,
                mask=real[:, None], other=0.0)
    acc = tl.dot(a, b, acc)

# the rest: G groups per dot, every group full
BKW:  tl.constexpr = 16 * G
STEP: tl.constexpr = G * KPD
ix   = tl.arange(0, BKW)
# lane ix holds k off+koff[ix]; KPD 8 zeroes each tail
koff = (ix // 16) * KPD + ix % 16
wr   = ix % 16 < KPD
for _ in tl.range(0, (klen - rbk) // STEP,
                 num_stages=NSTAGE):
    a = tl.load(ap, mask=wr[None, :], other=0.0)
    b = tl.load(bp, mask=wr[:, None], other=0.0)
    acc = tl.dot(a, b, acc)
    ap += STEP * ak
    bp += STEP * bk
\end{lstlisting}

Across the whole arm, 154,904 byte comparisons returned zero differences. The
load-bearing half of that is a sweep of the plan-parameter space rather than of
shapes: 7,868 parameter combinations over the nine modes at ten input draws
each. cuBLAS's own heuristic hands any one shape a small corner of a plan's
parameter space, so the sweep walks that space directly, and it was checked for
power against a deliberately broken kernel whose backwards split-K partials it
caught. Every input set was drawn twice over, once ordinarily and once
with the exponents spread across the dtype's usable range, since narrow
exponents hide a regrouping.

The descriptor path costs about 100\,$\mu$s of host time per call against about
20\,$\mu$s for the pointer launch, which is what builds the tensor descriptors.
Device time does not see it and a caller does, so it is visible on a GEMM
under roughly 100\,$\mu$s.

\section{The reduction order of attention}
\label{app:attention}

One attention kernel holds three reductions. The score matrix contracts $Q$
against $K$ over the head dimension, and the output contracts the probabilities
against $V$ over the key axis; both are matrix multiplications, and their order
is the object of Section~\ref{sec:theory-gemm}. Between them sits an
accumulation over the key axis that carries a scale factor, and that factor is
what separates attention from every reduction in
Section~\ref{sec:theory-reduction}.

\textbf{The recurrence.} Cut the keys of one query row into blocks
$\mathcal{B}_1,\dots,\mathcal{B}_B$ of width $w$, and write $s_t$ for the scaled
score of key $t$ and $v_t$ for its value row. The kernel carries three running
quantities in fp32: a maximum $m_j$, a denominator $\ell_j$ and an output row
$O_j$. At block $j$ it folds the block's own maximum into the running one and
rescales what it already holds by $\alpha_j$ before it adds. Writing
$m_j = \max(m_{j-1},\, \max_{t \in \mathcal{B}_j} s_t)$,
$\alpha_j = 2^{\,m_{j-1} - m_j}$ and $p_t = 2^{\,s_t - m_j}$, the two running
sums are $\ell_{j} = \alpha_j \ell_{j-1} \fpadd{fp32} \fpsum{fp32}_{t \in
\mathcal{B}_j} p_t$ and $O_{j} = \alpha_j O_{j-1} \fpadd{fp32}
\fpsum{fp32}_{t \in \mathcal{B}_j} p_t\, v_t$.
The base is two because the kernel folds $\log_2 e$ into the score scale and
calls the hardware's base-two exponential. The answer is $O_B / \ell_B$, taken
once at the end.

\textbf{Every block boundary is a rounding.} In exact arithmetic $\alpha_j$
undoes the previous normalisation, and the recurrence returns one value at every
$w$. In floating point $\alpha_j O_{j-1}$ is a multiply over the whole running
output row, so each boundary costs one rounding on every element of $O$ and one
on $\ell$, and a row of $N$ keys pays $\lceil N / w \rceil$ of them. The key
block width, \texttt{BLOCK\_N} in Triton, therefore sets how many roundings the
answer carries and where they fall.
Section~\ref{sec:theory-families} put \texttt{BLOCK\_K} on the other side of that
line: an accumulator carried across the turns of a tensor-core mainloop receives
the same instruction results in the same order however the turns are cut, so the
turn boundary is a speed knob there. Here it is a term of the arithmetic.

\textbf{Masking puts the query block into the answer as well.} Left unmasked,
the key walk runs from the first key to the last in steps of $w$, so the
boundaries are multiples of $w$ alone and the query block width,
\texttt{BLOCK\_M}, chooses only which rows share a kernel instance. Causal
masking splits the walk in two: a run of blocks lying entirely below the
diagonal, and one block straddling it, whose entries above the diagonal are
offset by $-10^{6}$ before the maximum is taken, so their probabilities
underflow to zero. Both bounds of that split are multiples of \texttt{BLOCK\_M},
so the query block width decides where the straddling block begins, how many
keys inside it are live, and how many rescales precede it. Two causal
configurations differing only in
\texttt{BLOCK\_M} accumulate over different boundaries, and two unmasked ones
accumulate over the same boundaries at any \texttt{BLOCK\_M}.

\textbf{What the layout still moves.} Two reductions run inside a block. The row
maximum selects one of its inputs and rounds nothing, so its dependence tree is
free. The row sum $\fpsum{fp32}_{t \in \mathcal{B}_j} p_t$ is a plain sum of $w$
values, which makes it the object of Section~\ref{sec:theory-reduction} exactly,
and its tree moves with the thread count and the lane layout in the way that
section describes.

\textbf{The probabilities are narrowed before the second matmul.} A tensor core
takes its operands at the input type, so $p_t$ is rounded from the fp32 it was
computed in down to fp16, bf16 or fp8 before it meets $V$. That rounding lands
on every probability of every block, and it sits between the two reductions the
kernel is composing, so the operand type is as much a part of the order as the
accumulator type is.

\textbf{The descriptor.} An attention descriptor carries the key block width,
the masking convention together with the query block width when masking is on,
the tree of the row sum inside a block, the type the probabilities are narrowed
to, the descriptors of the two matrix multiplications, and the place of the
final division.

\bibliographystyle{ACM-Reference-Format}
\bibliography{references}

@inproceedings{summers2021nondeterminism,
  author    = {Summers, Cecilia and Dinneen, Michael J.},
  title     = {Nondeterminism and Instability in Neural Network Optimization},
  booktitle = {Proceedings of the 38th International Conference on Machine Learning (ICML)},
  year      = {2021},
}

@misc{bhojanapalli2021reproducibility,
  author        = {Bhojanapalli, Srinadh and Wilber, Kimberly and Veit, Andreas and
                   Rawat, Ankit Singh and Kim, Seungyeon and Menon, Aditya Krishna and
                   Kumar, Sanjiv},
  title         = {On the Reproducibility of Neural Network Predictions},
  year          = {2021},
  eprint        = {2102.03349},
  archivePrefix = {arXiv},
  primaryClass  = {cs.LG},
}

@misc{shanmugavelu2024fpna,
  author        = {Shanmugavelu, Sanjif and others},
  title         = {Impacts of Floating-Point Non-Associativity on Reproducibility for
                   HPC and Deep Learning Applications},
  year          = {2024},
  eprint        = {2408.05148},
  archivePrefix = {arXiv},
  primaryClass  = {cs.DC},
}

@misc{llmnondeterminism2025,
  author        = {Yuan, Jiayi and Li, Hao and Ding, Xinheng and Xie, Wenya and
                   Li, Yu-Jhe and Zhao, Wentian and Wan, Kun and Shi, Jing and
                   Hu, Xia and Liu, Zirui},
  title         = {Understanding and Mitigating Numerical Sources of Nondeterminism in {LLM} Inference},
  year          = {2025},
  eprint        = {2506.09501},
  archivePrefix = {arXiv},
  primaryClass  = {cs.LG},
}

@misc{he2025defeating,
  author       = {He, Horace and {Thinking Machines Lab}},
  title        = {Defeating Nondeterminism in {LLM} Inference},
  year         = {2025},
  month        = sep,
  howpublished = {\url{https://thinkingmachines.ai/blog/defeating-nondeterminism-in-llm-inference/}},
  note         = {Thinking Machines Lab blog; accessed 2026-09-05},
}

@misc{repdl2025,
  author        = {Xie, Peichen and Zhang, Xian and Chen, Shuo},
  title         = {{RepDL}: Bit-level Reproducible Deep Learning Training and Inference},
  year          = {2025},
  eprint        = {2510.09180},
  archivePrefix = {arXiv},
  primaryClass  = {cs.LG},
}

@inproceedings{triton2019,
  author    = {Tillet, Philippe and Kung, H. T. and Cox, David},
  title     = {Triton: An Intermediate Language and Compiler for Tiled Neural Network Computations},
  booktitle = {Proceedings of the 3rd ACM SIGPLAN International Workshop on Machine
               Learning and Programming Languages (MAPL)},
  year      = {2019},
}

@inproceedings{linearlayouts2026,
  author    = {Zhou, Keren and Lezcano, Mario and Goucher, Adam and Rakhmati, Akhmed and
               Niu, Jeff and Lebar, Justin and Szczerbuk, Pawel and Bell, Peter and
               Tillet, Phil and Raoux, Thomas and Moudallal, Zahi},
  title     = {Linear Layouts: Robust Code Generation of Efficient Tensor Computation
               Using $\mathbb{F}_2$},
  booktitle = {Proceedings of the 31st ACM International Conference on Architectural
               Support for Programming Languages and Operating Systems (ASPLOS), Volume 1},
  year      = {2026},
  doi       = {10.1145/3760250.3762221},
}

@article{compcert2009,
  author  = {Leroy, Xavier},
  title   = {Formal Verification of a Realistic Compiler},
  journal = {Communications of the ACM},
  volume  = {52},
  number  = {7},
  pages   = {107--115},
  year    = {2009},
}

@inproceedings{alive2_2021,
  author    = {Lopes, Nuno P. and Lee, Juneyoung and Hur, Chung-Kil and
               Liu, Zhengyang and Regehr, John},
  title     = {Alive2: Bounded Translation Validation for {LLVM}},
  booktitle = {Proceedings of the 42nd ACM SIGPLAN Conference on Programming
               Language Design and Implementation (PLDI)},
  year      = {2021},
}

@inproceedings{alivefp2016,
  author    = {Menendez, David and Nagarakatte, Santosh and Gupta, Aarti},
  title     = {{Alive-FP}: Automated Verification of Floating Point Based
               Peephole Optimizations in {LLVM}},
  booktitle = {Static Analysis (SAS)},
  pages     = {317--337},
  year      = {2016},
  doi       = {10.1007/978-3-662-53413-7\_16},
}

@inproceedings{pnueli1998tv,
  author    = {Pnueli, Amir and Siegel, Michael and Singerman, Eli},
  title     = {Translation Validation},
  booktitle = {Tools and Algorithms for the Construction and Analysis of Systems (TACAS)},
  year      = {1998},
}

@misc{pytorch_determinism,
  author       = {{PyTorch Team}},
  title        = {Reproducibility and Deterministic Algorithms},
  howpublished = {\url{https://pytorch.org/docs/stable/notes/randomness.html}},
  year         = {2026},
  note         = {PyTorch documentation; accessed 2026-09-05},
}

@misc{cuda_prog_guide,
  author       = {{NVIDIA}},
  title        = {{CUDA} {C++} Programming Guide},
  howpublished = {\url{https://docs.nvidia.com/cuda/cuda-c-programming-guide/}},
  year         = {2026},
  note         = {Thread hierarchy: grid, thread block, warp; 1024 threads per
                  block; thread block clusters at compute capability 9.0},
}

@misc{hip_prog_model,
  author       = {{AMD}},
  title        = {{HIP} Programming Model},
  howpublished = {\url{https://rocm.docs.amd.com/projects/HIP/en/latest/understand/programming_model.html}},
  year         = {2026},
  note         = {Workgroup, wavefront and compute unit; wavefront 64 on CDNA,
                  32 or 64 on RDNA},
}

@article{lindholm2008tesla,
  author  = {Lindholm, Erik and Nickolls, John and Oberman, Stuart and Montrym, John},
  title   = {{NVIDIA} {Tesla}: A Unified Graphics and Computing Architecture},
  journal = {IEEE Micro},
  volume  = {28},
  number  = {2},
  pages   = {39--55},
  year    = {2008},
  doi     = {10.1109/MM.2008.31},
}

@article{nickolls2008cuda,
  author  = {Nickolls, John and Buck, Ian and Garland, Michael and Skadron, Kevin},
  title   = {Scalable Parallel Programming with {CUDA}},
  journal = {ACM Queue},
  volume  = {6},
  number  = {2},
  pages   = {40--53},
  year    = {2008},
  doi     = {10.1145/1365490.1365500},
}

@misc{volta_whitepaper,
  author       = {{NVIDIA}},
  title        = {{NVIDIA} {Tesla} {V100} {GPU} Architecture},
  howpublished = {\url{https://images.nvidia.com/content/volta-architecture/pdf/volta-architecture-whitepaper.pdf}},
  year         = {2017},
  note         = {WP-08608-001\_v1.1; independent thread scheduling, per-thread
                  program counters},
}

@misc{cdna3_whitepaper,
  author       = {{AMD}},
  title        = {{AMD} {CDNA} 3 Architecture},
  howpublished = {\url{https://www.amd.com/content/dam/amd/en/documents/instinct-tech-docs/white-papers/amd-cdna-3-white-paper.pdf}},
  year         = {2023},
  note         = {Compute unit: four SIMD units, 64\,KB local data share,
                  64-work-item wavefront},
}

@article{thakur2005collectives,
  author  = {Thakur, Rajeev and Rabenseifner, Rolf and Gropp, William},
  title   = {Optimization of Collective Communication Operations in {MPICH}},
  journal = {International Journal of High Performance Computing Applications},
  volume  = {19},
  number  = {1},
  pages   = {49--66},
  year    = {2005},
  doi     = {10.1177/1094342005051521},
}

@misc{blackwell_tuning,
  author       = {{NVIDIA}},
  title        = {{NVIDIA} {Blackwell} Tuning Guide},
  howpublished = {\url{https://docs.nvidia.com/cuda/blackwell-tuning-guide/index.html}},
  year         = {2026},
  note         = {Compute capability 10.0: 228\,KB shared memory per SM, 227\,KB
                  the most one thread block may claim, 32 thread blocks per SM},
}

@inproceedings{demmel2013reproducible,
  author    = {Demmel, James and Nguyen, Hong Diep},
  title     = {Fast Reproducible Floating-Point Summation},
  booktitle = {2013 IEEE 21st Symposium on Computer Arithmetic (ARITH)},
  pages     = {163--172},
  year      = {2013},
  doi       = {10.1109/ARITH.2013.9},
}

@article{collange2015reduction,
  author  = {Collange, Caroline and Defour, David and Graillat, Stef and
             Iakymchuk, Roman},
  title   = {Numerical Reproducibility for the Parallel Reduction on Multi- and
             Many-Core Architectures},
  journal = {Parallel Computing},
  volume  = {49},
  pages   = {83--97},
  year    = {2015},
  doi     = {10.1016/j.parco.2015.09.001},
}

@article{ahrens2020reproblas,
  author  = {Ahrens, Willow and Demmel, James and Nguyen, Hong Diep},
  title   = {Algorithms for Efficient Reproducible Floating Point Summation},
  journal = {ACM Transactions on Mathematical Software},
  volume  = {46},
  number  = {3},
  pages   = {1--49},
  year    = {2020},
  doi     = {10.1145/3389360},
}

@misc{mkl_cnr,
  author       = {{Intel}},
  title        = {Obtaining Numerically Reproducible Results
                  ({Conditional} {Numerical} {Reproducibility})},
  howpublished = {\url{https://www.intel.com/content/www/us/en/docs/onemkl/developer-guide-linux/2023-0/obtaining-numerically-reproducible-results.html}},
  year         = {2023},
  note         = {oneMKL Developer Guide for Linux; accessed 2026-09-05},
}

@misc{deepseekv4,
  author        = {{DeepSeek-AI}},
  title         = {{DeepSeek-V4}: Towards Highly Efficient Million-Token Context
                   Intelligence},
  year          = {2026},
  eprint        = {2606.19348},
  archivePrefix = {arXiv},
  primaryClass  = {cs.CL},
}

@article{ozaki2025,
  author  = {Uchino, Yuki and Ozaki, Katsuhisa and Imamura, Toshiyuki},
  title   = {Performance Enhancement of the {Ozaki} Scheme on Integer Matrix
             Multiplication Unit},
  journal = {The International Journal of High Performance Computing Applications},
  volume  = {39},
  number  = {3},
  pages   = {462--476},
  year    = {2025},
  doi     = {10.1177/10943420241313064},
}

@article{boldo2015fpcompile,
  author  = {Boldo, Sylvie and Jourdan, Jacques-Henri and Leroy, Xavier and
             Melquiond, Guillaume},
  title   = {Verified Compilation of Floating-Point Computations},
  journal = {Journal of Automated Reasoning},
  volume  = {54},
  number  = {2},
  pages   = {135--163},
  year    = {2015},
  doi     = {10.1007/s10817-014-9317-x},
}

@misc{fp16mismatch,
  author        = {Qi, Penghui and Liu, Zichen and Zhou, Xiangxin and
                   Pang, Tianyu and Du, Chao and Lee, Wee Sun and Lin, Min},
  title         = {Defeating the Training-Inference Mismatch via {FP16}},
  year          = {2025},
  eprint        = {2510.26788},
  archivePrefix = {arXiv},
  primaryClass  = {cs.LG},
}

@inproceedings{tvm2018,
  author    = {Chen, Tianqi and Moreau, Thierry and Jiang, Ziheng and Zheng, Lianmin and
               Yan, Eddie and Cowan, Meghan and Shen, Haichen and Wang, Leyuan and Hu, Yuwei and
               Ceze, Luis and Guestrin, Carlos and Krishnamurthy, Arvind},
  title     = {{TVM}: An Automated End-to-End Optimizing Compiler for Deep Learning},
  booktitle = {13th USENIX Symposium on Operating Systems Design and Implementation (OSDI)},
  pages     = {578--594},
  year      = {2018},
}

@inproceedings{ansor2020,
  author    = {Zheng, Lianmin and Jia, Chengfan and Sun, Minmin and Wu, Zhao and Yu, Cody Hao and
               Haj-Ali, Ameer and Wang, Yida and Yang, Jun and Zhuo, Danyang and Sen, Koushik and
               Gonzalez, Joseph E. and Stoica, Ion},
  title     = {Ansor: Generating High-Performance Tensor Programs for Deep Learning},
  booktitle = {14th USENIX Symposium on Operating Systems Design and Implementation (OSDI)},
  pages     = {863--879},
  year      = {2020},
}

@inproceedings{pytorch2_2024,
  author    = {Ansel, Jason and Yang, Edward and He, Horace and Gimelshein, Natalia and
               Jain, Animesh and Voznesensky, Michael and Bao, Bin and Bell, Peter and
               Berard, David and Burovski, Evgeni and Chauhan, Geeta and Chourdia, Anjali and
               Constable, Will and Desmaison, Alban and DeVito, Zachary and Ellison, Elias and
               Feng, Will and Gong, Jiong and Gschwind, Michael and Hirsh, Brian and
               Huang, Sherlock and Kalambarkar, Kshiteej and Kirsch, Laurent and Lazos, Michael and
               Lezcano, Mario and Liang, Yanbo and Liang, Jason and Lu, Yinghai and
               Luk, C. K. and Maher, Bert and Pan, Yunjie and Puhrsch, Christian and
               Reso, Matthias and Saroufim, Mark and Siraichi, Marcos Yukio and Susnea, Helen and
               Zhang, Shunting and Zhang, Michael and Zaharia, Matei and Chintala, Soumith},
  title     = {{PyTorch} 2: Faster Machine Learning Through Dynamic {Python} Bytecode
               Transformation and Graph Compilation},
  booktitle = {Proceedings of the 29th ACM International Conference on Architectural Support
               for Programming Languages and Operating Systems (ASPLOS), Volume 2},
  year      = {2024},
  doi       = {10.1145/3620665.3640366},
}

@inproceedings{flashattention2022,
  author    = {Dao, Tri and Fu, Daniel Y. and Ermon, Stefano and Rudra, Atri and R{\'e}, Christopher},
  title     = {{FlashAttention}: Fast and Memory-Efficient Exact Attention with {IO}-Awareness},
  booktitle = {Advances in Neural Information Processing Systems 35 (NeurIPS)},
  year      = {2022},
}

@misc{rocq_prover,
  author       = {{The Rocq Prover Team}},
  title        = {About The {Rocq} Prover},
  howpublished = {\url{https://rocq-prover.org/about}},
  year         = {2026},
  note         = {States that the Rocq Prover was formerly known as the Coq Proof
                  Assistant; accessed 2026-09-05},
}

@misc{huggingface_hub,
  author       = {{Hugging Face}},
  title        = {The Model Hub},
  howpublished = {\url{https://huggingface.co/docs/hub/models}},
  year         = {2026},
  note         = {Each model's \texttt{config.json} is the source of the layer dimensions
                  used here; the text-generation listing is the source of the model
                  selection. Read 2026-08-16},
}

@misc{tritonbench,
  author       = {{Meta PyTorch}},
  title        = {Tritonbench: A Collection of {PyTorch} Custom Operators with Example Inputs},
  howpublished = {\url{https://github.com/pytorch-labs/tritonbench}},
  year         = {2026},
  note         = {Accessed 2026-09-06},
}

@misc{flaggems,
  author       = {{FlagOpen}},
  title        = {{FlagGems}: A {Triton}-Powered Operator Library},
  howpublished = {\url{https://github.com/FlagOpen/FlagGems}},
  year         = {2026},
  note         = {Accessed 2026-09-06},
}

@misc{flashlinearattention,
  author       = {{FLA Organization}},
  title        = {Flash Linear Attention: Efficient {Triton} Implementations for Emerging
                  Model Architectures},
  howpublished = {\url{https://github.com/fla-org/flash-linear-attention}},
  year         = {2026},
  note         = {Accessed 2026-09-06},
}

@misc{torchao,
  author       = {{PyTorch}},
  title        = {torchao: {PyTorch} Architecture Optimization},
  howpublished = {\url{https://github.com/pytorch/ao}},
  year         = {2026},
  note         = {Accessed 2026-09-06},
}

@misc{triton_tutorials,
  author       = {{Triton Developers}},
  title        = {{Triton} Tutorials},
  howpublished = {\url{https://github.com/triton-lang/triton/tree/main/python/tutorials}},
  year         = {2026},
  note         = {The fused-attention tutorial is the source of the flash-attention kernel
                  graded here; accessed 2026-09-06},
}

@misc{cublas_docs,
  author       = {{NVIDIA}},
  title        = {{cuBLAS} Library Documentation},
  howpublished = {\url{https://docs.nvidia.com/cuda/cublas/}},
  year         = {2026},
  note         = {Results reproducibility: bitwise repeatability only on GPUs of the same
                  architecture with the same number of SMs, and not across toolkit versions},
}

@misc{cudnn2014,
  author        = {Chetlur, Sharan and Woolley, Cliff and Vandermersch, Philippe and
                   Cohen, Jonathan and Tran, John and Catanzaro, Bryan and Shelhamer, Evan},
  title         = {{cuDNN}: Efficient Primitives for Deep Learning},
  year          = {2014},
  eprint        = {1410.0759},
  archivePrefix = {arXiv},
  primaryClass  = {cs.NE},
}

@misc{rocblas_docs,
  author       = {{AMD}},
  title        = {{rocBLAS} Documentation},
  howpublished = {\url{https://rocm.docs.amd.com/projects/rocBLAS/en/latest/}},
  year         = {2026},
  note         = {The ROCm BLAS library, implemented in HIP and tuned for AMD GPUs},
}

@inproceedings{tilelang2026,
  author    = {Wang, Lei and Cheng, Yu and Shi, Yining and Mo, Zhiwen and Tang, Zhengju and
               Xie, Wenhao and Wu, Tong and Ma, Lingxiao and Xia, Yuqing and Xue, Jilong and
               Yang, Fan and Yang, Zhi},
  title     = {{TileLang}: Bridge Programmability and Performance in Modern Neural Kernels},
  booktitle = {International Conference on Learning Representations (ICLR)},
  year      = {2026},
  note      = {Oral},
}

@misc{cutile,
  author       = {{NVIDIA}},
  title        = {{cuTile} {Python} and the {CUDA} {Tile} {IR}},
  howpublished = {\url{https://docs.nvidia.com/cuda/tile-ir/latest/}},
  year         = {2025},
  note         = {A tile-level Python kernel language for CUDA, lowering through an
                  MLIR-based tile intermediate representation},
}

@misc{pallas,
  author       = {{JAX Developers}},
  title        = {Pallas: a {JAX} Kernel Language},
  howpublished = {\url{https://docs.jax.dev/en/latest/pallas/index.html}},
  year         = {2026},
  note         = {Lowers to Mosaic on TPU and to Triton on GPU},
}

@misc{nki,
  author       = {{Amazon Web Services}},
  title        = {{Neuron} {Kernel} {Interface} ({NKI})},
  howpublished = {\url{https://awsdocs-neuron.readthedocs-hosted.com/en/latest/nki/}},
  year         = {2026},
  note         = {Tile-level Python kernel programming for AWS Trainium and Inferentia},
}

@inproceedings{tpu2017,
  author    = {Jouppi, Norman P. and Young, Cliff and Patil, Nishant and Patterson, David and
               Agrawal, Gaurav and Bajwa, Raminder and Bates, Sarah and Bhatia, Suresh and
               Boden, Nan and Borchers, Al and others},
  title     = {In-Datacenter Performance Analysis of a Tensor Processing Unit},
  booktitle = {Proceedings of the 44th Annual International Symposium on Computer
               Architecture (ISCA)},
  pages     = {1--12},
  year      = {2017},
  doi       = {10.1145/3079856.3080246},
}

@misc{trainium,
  author       = {{Amazon Web Services}},
  title        = {{AWS} {Trainium} and {Inferentia}},
  howpublished = {\url{https://aws.amazon.com/ai/machine-learning/trainium/}},
  year         = {2026},
  note         = {Amazon's training and inference accelerators},
}

@article{verifiedsched2024,
  author  = {Yang, Ziteng and Shirako, Jun and Sarkar, Vivek},
  title   = {Fully Verified Instruction Scheduling},
  journal = {Proceedings of the ACM on Programming Languages},
  volume  = {8},
  number  = {OOPSLA2},
  pages   = {791--816},
  year    = {2024},
  doi     = {10.1145/3689739},
}

@article{verifiedtensor2024,
  author  = {Liu, Amanda and Bernstein, Gilbert and Chlipala, Adam and
             Ragan-Kelley, Jonathan},
  title   = {A Verified Compiler for a Functional Tensor Language},
  journal = {Proceedings of the ACM on Programming Languages},
  volume  = {8},
  number  = {PLDI},
  pages   = {320--342},
  year    = {2024},
  doi     = {10.1145/3656390},
}

@inproceedings{ptxmemmodel2019,
  author    = {Lustig, Daniel and Sahasrabuddhe, Sameer and Giroux, Olivier},
  title     = {A Formal Analysis of the {NVIDIA} {PTX} Memory Consistency Model},
  booktitle = {Proceedings of the 24th ACM International Conference on
               Architectural Support for Programming Languages and Operating
               Systems (ASPLOS)},
  pages     = {257--270},
  year      = {2019},
  doi       = {10.1145/3297858.3304043},
}

@inproceedings{cudaaucoq2019,
  author    = {Ferrell, Benjamin and Duan, Jun and Hamlen, Kevin W.},
  title     = {{CUDA} au {Coq}: A Framework for Machine-validating {GPU}
               Assembly Programs},
  booktitle = {Design, Automation and Test in Europe Conference (DATE)},
  pages     = {474--479},
  year      = {2019},
  doi       = {10.23919/DATE.2019.8715160},
}

@inproceedings{mlirtv2022,
  author    = {Bang, Seongwon and Nam, Seunghyeon and Chun, Inwhan and
               Jhoo, Ho Young and Lee, Juneyoung},
  title     = {{SMT}-Based Translation Validation for Machine Learning Compiler},
  booktitle = {Computer Aided Verification (CAV), Part II},
  pages     = {386--407},
  year      = {2022},
  doi       = {10.1007/978-3-031-13188-2\_19},
}

@inproceedings{csmith2011,
  author    = {Yang, Xuejun and Chen, Yang and Eide, Eric and Regehr, John},
  title     = {Finding and Understanding Bugs in {C} Compilers},
  booktitle = {Proceedings of the 32nd ACM SIGPLAN Conference on Programming Language
               Design and Implementation (PLDI)},
  pages     = {283--294},
  year      = {2011},
  doi       = {10.1145/1993498.1993532},
}

@inproceedings{mlirsmith2023,
  author    = {Wang, Haoyu and Chen, Junjie and Xie, Chuyue and Liu, Shuang and
               Wang, Zan and Shen, Qingchao and Zhao, Yingquan},
  title     = {{MLIRSmith}: Random Program Generation for Fuzzing {MLIR} Compiler
               Infrastructure},
  booktitle = {Proceedings of the 38th IEEE/ACM International Conference on Automated
               Software Engineering (ASE)},
  pages     = {1555--1566},
  year      = {2023},
  doi       = {10.1109/ASE56229.2023.00120},
}

@inproceedings{flit2017,
  author    = {Sawaya, Geoffrey and Bentley, Michael and Briggs, Ian and
               Gopalakrishnan, Ganesh and Ahn, Dong H.},
  title     = {{FLiT}: Cross-Platform Floating-Point Result-Consistency Tester and
               Workload},
  booktitle = {2017 IEEE International Symposium on Workload Characterization
               (IISWC)},
  pages     = {229--238},
  year      = {2017},
  publisher = {IEEE},
  doi       = {10.1109/IISWC.2017.8167780},
}

@inproceedings{varity2020,
  author    = {Laguna, Ignacio},
  title     = {{Varity}: Quantifying Floating-Point Variations in {HPC} Systems
               Through Randomized Testing},
  booktitle = {2020 IEEE International Parallel and Distributed Processing
               Symposium (IPDPS)},
  pages     = {622--633},
  year      = {2020},
  publisher = {IEEE},
  doi       = {10.1109/IPDPS47924.2020.00070},
}

@incollection{blelloch_maggs_parallel,
  author    = {Blelloch, Guy E. and Maggs, Bruce M.},
  title     = {Parallel Algorithms},
  booktitle = {Algorithms and Theory of Computation Handbook},
  editor    = {Atallah, Mikhail J.},
  edition   = {2},
  publisher = {Chapman and Hall/CRC},
  year      = {2010},
  pages     = {25-1--25-43},
  doi       = {10.1201/9781584888215-c25}
}

@book{clrs,
  author    = {Cormen, Thomas H. and Leiserson, Charles E. and Rivest, Ronald L. and Stein, Clifford},
  title     = {Introduction to Algorithms},
  edition   = {4},
  publisher = {MIT Press},
  address   = {Cambridge, MA},
  year      = {2022},
  isbn      = {9780262046305}
}

@article{kuck_muraoka1974,
  author    = {Kuck, David J. and Muraoka, Yoichi},
  title     = {Bounds on the Parallel Evaluation of Arithmetic Expressions Using Associativity and Commutativity},
  journal   = {Acta Informatica},
  volume    = {3},
  number    = {3},
  pages     = {203--216},
  year      = {1974},
  publisher = {Springer},
  doi       = {10.1007/BF00288634}
}

@misc{mmasim,
  author       = {Xie, Peichen and Xu, Shuotao and Wang, Yang and Yang, Fan and Yang, Mao},
  title        = {Bit-Accurate Modeling of {GPU} Matrix Multiply-Accumulate Units:
                  Demystifying Numerical Discrepancy and Accuracy},
  year         = {2025},
  eprint       = {2511.10909},
  archivePrefix = {arXiv},
  primaryClass = {cs.AR},
  note         = {Bit-accurate models of every MMA instruction on ten GPU architectures,
                  NVIDIA Volta through Blackwell and AMD CDNA1 through CDNA3;
                  open source as MMA-Sim}
}

@misc{tcmodels,
  author       = {Khattak, Faizan A. and Mikaitis, Mantas},
  title        = {Accurate Models of {NVIDIA} Tensor Cores},
  year         = {2025},
  eprint       = {2512.07004},
  archivePrefix = {arXiv},
  primaryClass = {cs.MS},
  note         = {Bit-accurate inner-product models for V100, A100, H100 and B200 at
                  8-, 16- and 19-bit inputs, verified against the hardware}
}

@article{agarwal1995,
  author  = {Agarwal, R. C. and Balle, S. M. and Gustavson, F. G. and Joshi, M. and Palkar, P.},
  title   = {A Three-Dimensional Approach to Parallel Matrix Multiplication},
  journal = {IBM Journal of Research and Development},
  volume  = {39},
  number  = {5},
  pages   = {575--582},
  year    = {1995},
  month   = sep,
  doi     = {10.1147/rd.395.0575}
}

@article{blas2_1988,
  author  = {Dongarra, Jack J. and Du Croz, Jeremy and Hammarling, Sven and Hanson, Richard J.},
  title   = {An Extended Set of {FORTRAN} Basic Linear Algebra Subprograms},
  journal = {ACM Transactions on Mathematical Software},
  volume  = {14},
  number  = {1},
  pages   = {1--17},
  year    = {1988},
  month   = mar,
  doi     = {10.1145/42288.42291}
}



\end{document}